\documentclass{aa}
\usepackage{etoolbox}
\usepackage{graphicx}
\usepackage{txfonts}
\usepackage{float}
\usepackage{hyperref}
\hypersetup{
     colorlinks = true,
     linkcolor = blue,
     anchorcolor = blue,
     citecolor = blue,
     filecolor = blue,
     urlcolor = blue
     }
\usepackage{pdflscape} 
\usepackage{afterpage} 
\usepackage{tabularx}
\usepackage{breqn}
\usepackage{geometry}
\usepackage{lipsum}

\makeatletter
\renewcommand*\aa@pageof{, page \thepage{} of \pageref*{LastPage}}
\usepackage{caption}
\usepackage{multirow}
\usepackage{lscape}
\usepackage{orcidlink}
\usepackage{float}
\newcommand*{\bba}{$^{\scriptstyle 3\mathrm{D}}${\rm {Barolo}}}
\defcitealias{paper_galaxyhalo}{MP25}
\defcitealias{santos2020}{SS20}
\newcommand*{\hi}{H\,{\sc i}\xspace}
\begin{document} 

   \title{
   Dark matter haloes from dwarf to massive galaxies: No systematic inner-density tension with $\Lambda$CDM hydrodynamical simulations} 
   \titlerunning{Dark matter from dwarf to massive galaxies: No inner-density tension with simulations}
\authorrunning{Mancera Pi\~na et al.}

   \author{Pavel E. Mancera Pi\~na\inst{1}\fnmsep\thanks{\email{pavel@strw.leidenuniv.nl}}\,\orcidlink{0000-0001-5175-939X}, 
Justin I. Read\inst{2}\,\orcidlink{0000-0002-1164-9302},
Jorge Sarrato-Alós\inst{3,4}\,\orcidlink{0009-0008-6119-1644},
Claudia Muni\inst{5}\,\orcidlink{0009-0005-7216-4922}
          }
   \institute{Leiden Observatory, Leiden University, P.O. Box 9513, 2300 RA, Leiden, The Netherlands 
   \and Department of Physics, University of Surrey, Guildford, Surrey GU2 7XH, United Kingdom
   \and Instituto de Astrofísica de Canarias (IAC), Calle Vía Láctea s/n, E-38205 La Laguna, Tenerife, Spain
   \and Universidad de La Laguna, Avda. Astrofísico Fco. Sánchez s/n, E-38206 La Laguna, Tenerife, Spain
   \and \mbox{The Oskar Klein Centre, Department of Physics, Stockholm University, Albanova University Center, 106 91 Stockholm, Sweden}
    }

   \date{}

 
  \abstract{
Two of the most prominent small-scale challenges to the standard cold dark matter (CDM) paradigm are the cusp--core and diversity-of-rotation-curves problems. The former concerns the shallow inner dark matter density profiles inferred for many galaxies compared with the cusps predicted by collisionless CDM simulations, while the latter concerns the wider range of inner dark matter densities and rotation curve shapes inferred observationally than hydrodynamical simulations traditionally reproduce. Robust observational constraints on dark matter core sizes and inner halo densities are therefore essential for testing both the nature of dark matter and the impact of galaxy formation processes.

\looseness=-1
In this work, we analyse the inner dark matter distribution of a curated sample of 48 gas-rich disc galaxies and eight Milky Way gas-poor satellites, spanning six orders of magnitude in stellar mass ($M_\ast$). We find substantial galaxy-to-galaxy scatter in dark matter core sizes and degrees of coreness, with both cuspy and cored haloes occurring over a broad range in $M_\ast$. The inferred cores are energetically consistent with stellar feedback, requiring modest supernova energy coupling efficiencies typically of the order of $0.1{-}1\%$. Comparisons with the NIHAO, FIRE-2, and EDGE hydrodynamical simulations reveal broad agreement in the inner dark matter densities and logarithmic slopes of observed and simulated galaxies, both at homologous halo radii and at fixed physical scales. The main residual differences concern the steep slopes of some massive simulated galaxies (arguably due to baryonic contraction) and differences in the stellar-to-halo mass relation. Applying a diagnostic of rotation-curve diversity used in previous work, we find that the extreme discrepancies with simulations are absent from our curated sample and are largely attributable to uncertain kinematics or baryonic mass distributions.
Within the observational and numerical scope of our analysis, we find no evidence of a systematic inner-density tension between our galaxy sample and current $\Lambda$CDM hydrodynamical simulations. Together with the modest energetic requirements for core formation, this substantially alleviates the cusp-core and diversity-of-rotation-curves problems.}


   \maketitle
%
\section{Introduction}
\label{sec:intro}
A large number of dwarf galaxies with stellar masses of $M_\ast\,{\lesssim}\,10^9\, M_\odot$ appear to have cored dark matter density profiles, with an inner logarithmic slope of $\alpha\,{\equiv}\, d\ln\rho_{\rm DM}/d\ln r\,{\approx}\,0$, according to dynamical modelling of their stellar or gas kinematics
(e.g. \citealt{moore1994,burkert1995,donato2004,gilmore2007,deblok2001,oh2015,salucci2019}). By contrast, collisionless cold dark matter (CDM) simulations produce dark matter haloes with central density cusps, corresponding to $\alpha\,{\approx}\,{-}1$ \citep{nfw,navarro2010,bullock2017}.

Understanding the origin of dark matter cores remains a subject of active debate, involving the delicate interplay between dark matter and baryons \citep{collins_feedback,sales_review_dwarfs}. While observational caveats in the interpretation of galaxy kinematics remain (e.g. \citealt{battaglia2008,battaglia2013,readAD,genina2018,read2021,downing2023,chase2026,dado2026}), theoretical studies within CDM have shown that cored dark matter distributions can arise in low-mass galaxies following feedback episodes that modify their initially cuspy haloes. Specifically, energy released by supernovae (SNe) can redistribute gas and generate rapid fluctuations in the gravitational potential. These fluctuations dynamically heat the dark matter and displace it outwards from the centre, thereby producing a core (e.g. \citealt{navarro_cores,read2005,pontzen2012,penarrubia2012,maxwell2015,gonzalez2016,burger2021}). Feedback from active galactic nuclei (AGN) may also contribute to this process \citep{martizzi2013,koudmani2025}. The efficiency of the cusp--core transformation appears to depend on the stellar-to-halo mass ratio \citep{dicintio2014,dutton2016,sales_review_dwarfs}, as well as on the timing of star formation. Core formation may be more efficient in galaxies with prolonged star formation histories (SFHs; \citealt{onorbe2015,read2019,bouche2022,azartash2024}, but see also \citealt{benitezllambay19,ciocan2026}), and in those that formed a larger fraction of their stellar mass after re-ionisation \citep{muni2025,sarrato2026}.

However, the debate is far from settled. While some simulations do not promote core formation (e.g. \citealt{lovell2018,revaz2018,bose2019,jahn2023}), different studies have found that cores generically form in simulations that resolve the cold and hot gas phases at the centres of dwarf galaxies and model galaxies massive enough to form stars after re-ionisation (e.g. \citealt{pontzen2012,dicintio2014,orkney2021,muni2025,sarrato2026}). Yet simulations that form dark matter cores in dwarf galaxies disagree on detailed sizes, inner density slopes, mass thresholds for core formation, and scatter \citep{chan2015,coreNFW,fitts2017,coreEinasto,jahn2023,jackson2025}. This is due to different numerical resolutions and sub-grid physics choices, all of which can impact core formation. Furthermore, studies have suggested that cored dark matter distributions may also be present in Milky Way-sized systems \citep{binney2001_MW,donato2004}, and some simulations predict that such cores can develop through stellar or AGN feedback \citep{chan2015,coreEinasto,maccio2020}.

These aspects are commonly discussed in the context of the cusp–core and diversity-of-rotation-curves problems. The former concerns the apparent discrepancy between the cusps predicted by collisionless CDM simulations and the shallower inner profiles inferred for many galaxies. The latter concerns the wide galaxy-to-galaxy variation in rotation curve shapes and inferred inner dark matter distributions, which has traditionally been larger in observations than in simulations \citep{oman2015,santos2020,sales_review_dwarfs}. Tensions have also been reported on larger scales, with some modern cosmological hydrodynamical simulations producing systems that are more dark matter dominated and exhibit stellar-to-halo mass relations (SHMRs) with less scatter than inferred observationally \citep{marasco_DMinsim,romeo2020_instabilities,paper_galaxyhalo}, as well as low-mass dwarfs that are thicker and less rotationally supported than observed \citep{benavides2025,benavides2026}.

Characterising both the inner dark matter distribution and its connection to the global halo is therefore key to constraining the nature of dark matter and its role in galaxy evolution. Traditionally, studies using neutral atomic hydrogen (\hi) rotation curves have described the dark matter distribution using halo profiles that, while phenomenological, lack a direct cosmological basis. These include the pseudo-isothermal \citep{begeman} and Burkert \citep{burkert1995} profiles (e.g. \citealt{deblok2001,oh2015}). However, cosmologically motivated halo profiles that attempt to capture the expected response of dark matter to baryonic processes are now available (e.g. \citealt{dicintio2014_haloprofile,coreNFW,coreEinasto}). Recent techniques have also made it possible to incorporate realistic gas disc thicknesses into rotation curve decompositions. This can significantly affect the recovered dark matter parameters of the smallest gas-rich dwarfs, which are crucial for testing dark matter and cosmological models \citep{paper_massmodels,paper_galaxyhalo}.

These advances, together with increasingly refined kinematic and dynamical models for both gas-rich and gas-poor galaxies (e.g. \citealt{iorio,alvarez2020,enrico_radialmotions,battaglia2022,paper_galaxyhalo}), motivated us to explore the inner dark matter densities and core properties of nearby galaxies. We have done so by analysing dynamical models of galaxies spanning nearly six orders of magnitude in $M_\ast$. We measured their core sizes, degrees of coreness, inner dark matter densities, and logarithmic slopes, and investigated how these properties depend on stellar and halo mass. We also examined whether the inferred cores are energetically compatible with stellar feedback and compared the recovered halo properties directly with predictions from $\Lambda$CDM hydrodynamical simulations. This allowed us to assess whether the observed diversity of halo densities constitutes a systematic tension with current simulations. 

The rest of this paper is organised as follows. In Section~\ref{sec:sample}, we introduce the galaxy sample and the mass models used in our analysis. In Section~\ref{sec:results_discussion}, we present and discuss our main results: the inferred dark matter core sizes and degree of coreness in our galaxy sample (Section~\ref{sec:cores}), the energetics of the cusp-core transformation (Section~\ref{sec:energy}), the dark matter densities and slopes of our data compared to expectations from hydrodynamical simulations (Sections~\ref{sec:simulations} and \ref{sec:comparison}), and caveats related to our analysis (Section~\ref{sec:caveats}). Finally, we present our summary and conclusions in Section~\ref{sec:conclusions}. Throughout this work, we adopt a $\Lambda$CDM cosmology with $\Omega_{\rm m} = 0.3$, $\Omega_{\Lambda} = 0.7$ and $H_0 = 70~\rm{km\,s^{-1}\,Mpc^{-1}}$.

\section{Sample overview and mass models}
\label{sec:sample}

\subsection{Gas-rich disc galaxies}
In this work, we first rely on the curated sample of gas-rich disc galaxies compiled by \citet[][hereafter \citetalias{paper_galaxyhalo}]{paper_galaxyhalo}. As detailed in \citetalias{paper_galaxyhalo}, the galaxies are selected to have $i)$ high-resolution public \hi data showing regular velocity fields and position-velocity slices without signs of strong non-circular motions; $ii)$ available near-infrared photometry (at $3.6\,\mu\rm{m}$ or $1.65\,\mu\rm{m}$, see \citetalias{paper_galaxyhalo}), including 2D bulge-disc decomposition for the massive galaxies \citep{salo2015}; $iii)$ inclination angles between $35^\circ$ and $80^\circ$; and $iv)$ accurate distances from the tip of the red giant branch, SNe, or Cepheids. After excluding the galaxy DDO 210 (the galaxy with the shortest radial coverage), the sample includes 48 galaxies spanning the stellar and baryonic mass range $2\,{\times}\,10^6\, {\lesssim}\, M_\ast/M_\odot\, {\lesssim}\, 2\,{\times}\,10^{11}$ and $3\,{\times}\,10^7\, {\lesssim}\, M_{\rm bar}/M_\odot\, {\lesssim}\, 3\,{\times}\,10^{11}$, respectively, with $M_{\rm bar}\,{=}\, M_\ast\,{+}\, M_{\rm gas}$, and $M_{\rm gas}$ the sum of the \hi mass and, when present, molecular gas (H$_2$) mass, both corrected for helium.

Crucially, all the galaxies have robust kinematic measurements derived by \cite{iorio}, \cite{enrico_radialmotions}, and \citetalias{paper_galaxyhalo} from accurate modelling of their \hi data cubes using the software \bba, which mitigates beam smearing effects \citep{barolo}. The sample spans rotational velocities in the range $20\,{\lesssim}\ V_{\rm rot}/\mathrm{km\,s^{-1}}\,{\lesssim}\,300$. Moreover, the rotation curves are corrected for turbulent motions through the asymmetric drift correction (e.g. \citealt{oh2015,readAD,paperIBFR}), following the procedure detailed in \cite{iorio}. All of the above make this a golden sample for dynamical studies such as mass modelling through rotation curve decomposition.

\looseness=-1
The technique of rotation curve decomposition consists of fitting the circular speed of disc galaxies ($V_{\rm circ}$, i.e. the rotation curve after correcting for asymmetric drift) with a model velocity $V_{\rm circ, mod}$ that incorporates their baryonic and dark matter contributions such that
\begin{equation}
\label{eq:massmodel}
V_{\rm circ,mod}^2
=
\Upsilon_{\rm d}V_{\rm d}^2
+\Upsilon_{\rm b}V_{\rm b}^2
+V_{\rm HI}^2
+V_{\rm H_2}^2
+V_{\rm DM}^2~.
\end{equation}
Here, $V_i^2(R)\equiv R\,\partial\Phi_i/\partial R$ denotes the contribution of component $i$ to $V_{\rm circ, mod}^2$ and may be negative where that component exerts an outward radial force. The stellar-disc and bulge contributions, $V_{\rm d}^2$ and $V_{\rm b}^2$, are calculated for unit mass-to-light ratio and normalised by $\Upsilon_{\rm d}$ and $\Upsilon_{\rm b}$, respectively. The radial dependence of the baryonic contributions is set by the corresponding observed light or gas surface-density distributions.

The dark matter contribution ($V_{\rm DM}(R)\, {=}\, \sqrt{G\, M_{\rm DM}/R}\,$) is often parametrised with functional forms such as the NFW profile \citep{nfw}, which has a density profile given by
\begin{equation}
    \rho_{\rm NFW}(r) = \dfrac{4\,\rho_{\rm s}}{(r/r_{\rm s})\,(1 + r/r_{\rm s})^2}~,
\end{equation}
and the enclosed mass profile
\begin{multline}
    M_{\rm NFW}(<r) = \dfrac{M_{200}}{\ln(1+c_{200}) - \dfrac{c_{200}}{1+c_{200}}} \\ 
    \times\, \left[\ln\left( 1 + \dfrac{r}{r_{\rm s}}\right) - \dfrac{r}{r_{\rm s}} \left( 1 + \dfrac{r}{r_{\rm s}}\right)^{-1} \right]~,
\end{multline}
with $r = \sqrt{R^2 + z^2}$ the spherical radius, $r_{\rm s}$ a scale radius, $\rho_{\rm s}$ the volume density at $r_{\rm s}$, $M_{200}$ the halo mass within the radius $R_{200}$ (where the average density is 200 times the critical density of the universe), and $c_{200} = R_{200}/r_{\rm s}$ the concentration parameter.  

However, in this work, we rely on the more flexible \textsc{coreNFW} profile \citep{coreNFW,readAD}, with density profile
\begin{equation}
\label{eq:corenfw_1}
    \rho_{\rm\textsc{cNFW}}(r) = f^n\, \rho_{\rm NFW}(r) + \dfrac{n\,f^{n-1}\,(1-f^2)}{4\,\pi\, r^2\, r_{\rm c}} M_{\rm NFW}(r)~,
\end{equation}
and an enclosed mass profile given by
\begin{equation}
\label{eq:corenfw_1_mass}
    M_{\rm\textsc{cNFW}}(r) = M_{\rm NFW}(r)\,f^n~,
\end{equation}
where $f\,{=}\,\tanh(r/r_{\rm c})$ generates a core of size $r_{\rm c}$, and $n$ is a parameter that regulates the cusp--core transition or degree of coreness. In the limit $n\,{=}\,0$ the profile reduces to an NFW halo, while $n\,{=}\,1$ produces a fully cored profile. 

\begin{figure}
    \centering
    \includegraphics[width=1\linewidth]{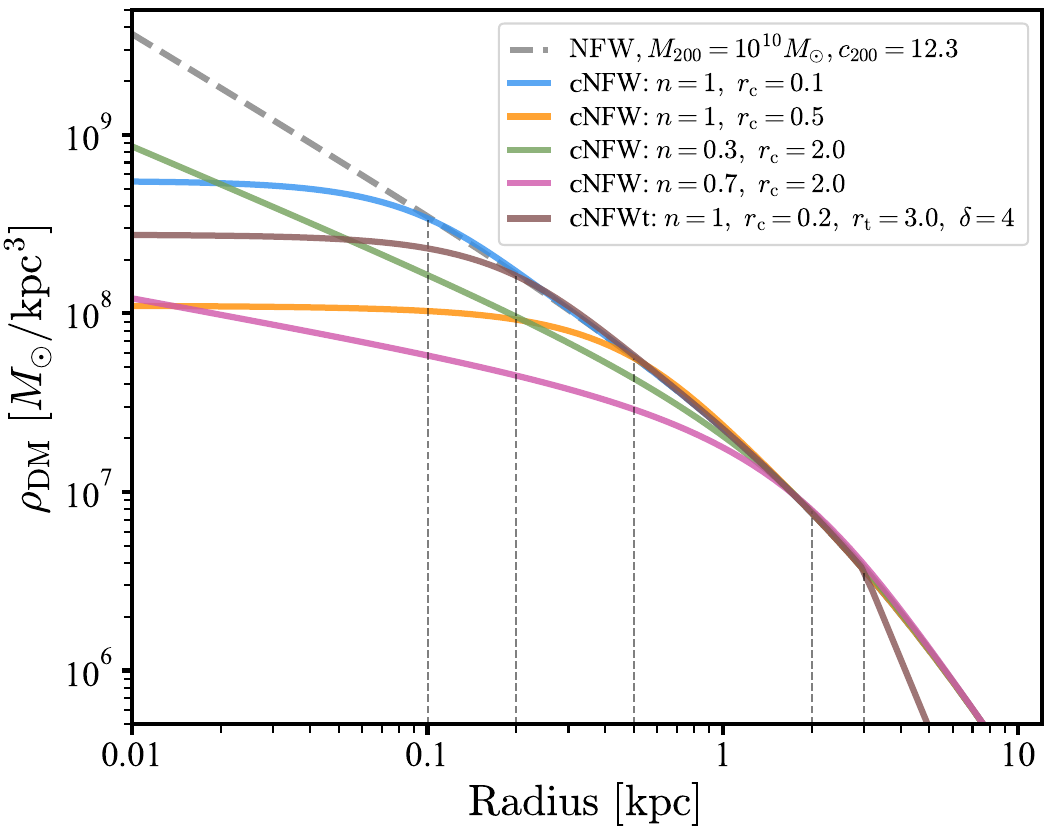}
    \caption{Illustrative \textsc{coreNFW} and \textsc{coreNFWtides} dark matter density profiles (solid coloured curves) compared to an NFW halo (dashed grey curve). All the profiles have the same halo mass and concentration. Dashed thin vertical lines indicate different values for the $r_{\rm c}$ and $r_{\rm t}$ parameters.
    }
    \label{fig:example_cNFW}
\end{figure}

\looseness=-1
It is instructive to visualise the effect of $n$ and $r_{\rm c}$ on the \textsc{coreNFW} profile. In Figure~\ref{fig:example_cNFW}, we show a set of density profiles that vary those two parameters, compared to a typical NFW profile (all the profiles have the same mass and concentration). The haloes with $n\,{=}\,1$ are fully cored and depart clearly from the NFW halo up to $r_{\rm c}$. The haloes with a fixed $r_{\rm c}$ and different $n$ also follow the NFW profile beyond $r_{\rm c}$, and in the inner regions they have intermediate slopes between the NFW and the fully cored profiles. This highlights a key feature of the \textsc{coreNFW} profile: $r_{\rm c}$ visually matches the radius at which the profiles clearly differ from NFW. We remind the reader that \textsc{coreNFW} haloes preserve the NFW mass, displacing the mass to regions slightly beyond $r_{\rm c}$. The displacement is more gentle and confined than in other feedback-modified profiles such as the \textsc{coreEinasto} \citep{coreEinasto} or the cored profile from \cite{penarrubia2012}, and requires lower supernova energies to unbind the dark matter halo and form sizeable cores (see also \citealt{maxwell2015,vogl2025}).

In general, a mass model such as the one described above has six free parameters: $\Upsilon_{\rm d}$, $\Upsilon_{\rm b}$, $M_{200}$, $c_{200}$, $r_{\rm c}$, and $n$. In practice, it is convenient to parametrise $r_{\rm c}$ as $r_{\rm c}\,{=}\,\eta\, R_{\rm eff}$, with $R_{\rm eff}$ the half-light radius and $\eta$ a fitting parameter. Ideally, both $n$ and $\eta$ would be treated as free parameters. However, \cite{read2017} and \citetalias{paper_galaxyhalo} found that often the two parameters are unconstrained when fitted simultaneously at the typical resolution of \hi data. Therefore, in this study, we use two sets of mass models: one that keeps $n$ fixed and $\eta$ free, and one that keeps $\eta$ fixed and $n$ free.

\subsubsection{Literature mass models with $n\,{=}\,1$ and free $\eta$}
\label{sec:original_models}
The first mass models are taken directly from \citetalias{paper_galaxyhalo}. Here we summarise their methods and refer readers to that reference for full details. For $\Upsilon_{\rm d}$ and $\Upsilon_{\rm b}$, \citetalias{paper_galaxyhalo} considered Gaussian priors centred on luminosity$-\Upsilon$ relations informed by spectral energy distribution (SED) fitting (see their Appendix B and \citealt{marasco_mstar}). For $M_{200}$, the flat prior $6 < \log(M_{200}/M_\odot) < 14$ was used. For $c_{200}$ a Gaussian prior on the  $c_{200}{-}M_{200}$ relation of \cite{diemer2019} was adopted. Moreover, $n$ was fixed to 1 (i.e. full core transformation), but $\eta$ was left free with a flat prior $\log(0.1)\, {\leq}\, \log(\eta)\, {\leq}\, \log(3.75)$, motivated by energy arguments \citep{coreNFW,readAD}. The minimisation was performed using a standard $\chi^2$ likelihood and the nested-sampling software \texttt{dynesty} \citep{dynesty}. Crucially, the scale height of the gaseous discs (which in turn affects $V_{\rm{HI}}$ and $V_{\rm{H_2}}$) was simultaneously and self-consistently derived with the global mass model using \textsc{galpynamics} \citep{iorio_phd} as implemented in \citet{paper_massmodels} and \citetalias{paper_galaxyhalo}.

\subsubsection{Additional mass models with fixed $\eta\,{=}\,1.5$ and free $n$}
\label{sec:new_models}
In this work, we explore an additional set of mass models. The methodology to obtain the new mass models is the same as in \citetalias{paper_galaxyhalo}, except for one difference. Now, $n$ is treated as a free parameter with a flat prior over $0 \leq n \leq 1$, and $\eta$ is fixed at $1.5$, such that $r_{\rm c}\,{=}\,1.5\, R_{\rm eff}$. The value of $\eta$ is chosen based on observational results from \citetalias{paper_galaxyhalo} (which we show below) and \cite{read2017}, as well as on theoretical studies using simulated galaxies \citep{coreNFW,readAD}. 
Our new approach yields mass models that closely reproduce the observed circular speed profiles of our galaxies. The rotation curve decomposition plots and posterior distributions can be found at \href{https://www.dropbox.com/scl/fo/8d7sh6o8cd9us6rfd995x/AHxoeqKENB3aipxIEjklgl4?rlkey=uk7uctvpo6msx10z7shs115f5&st=kxx2lrt3&dl=0}{this link}. 
Both sets of mass models yield comparable fits and best-fit parameters consistent within their uncertainties. The only galaxy with significant differences larger than the reported uncertainties is WLM, for which the mass models with free $n$ recover a higher $\log(M_{200}/M_\odot)$ and a lower $\log(c_{200})$ by 0.94 dex and 0.6 dex. The new mass models described in this section are the primary set we use throughout the text.

\subsection{Gas-poor galaxies}
We complement our gas-rich discs with a sample of eight dwarf spheroidal Milky Way satellites studied by \cite{read2019}. These galaxies have $3\,{\times}\,10^5 \lesssim M_\ast/M_\odot \lesssim 4.3\,{\times}\,10^7$ and negligible cold-gas reservoirs. The stellar kinematics of the galaxies have been modelled in \cite{alvarez2020} by solving the Jeans equations \citep{binney} with the software \textsc{GravSphere} \citep{read2017_grav,read2018,collins2021} and adopting a \textsc{coreNFWtides} dark matter halo.

The \textsc{coreNFWtides} halo \citep{read2018} is appropriate for satellite galaxies that may have experienced dark matter stripping by their environment, resulting in a steeper outer dark matter density. We note that, as for the \textsc{coreNFW} profile, $M_{200}$ is preserved for the \textsc{coreNFWtides} halo relative to an NFW halo, and should be interpreted as the satellites' pre-infall halo mass.

The density profile of the \textsc{coreNFWtides} halo is given by
\begin{equation}
\rho_{\rm cNFWt}(r) =
\begin{cases}
\rho_{\rm cNFW}(r),
& r < r_{\rm t}, \\[6pt]
\rho_{\rm cNFW}(r_{\rm t})
\left(\dfrac{r}{r_{\rm t}}\right)^{-\delta},
& r > r_{\rm t}~,
\end{cases}
\end{equation}
with a cumulative mass profile 

\begin{equation}
M_{\rm cNFWt}(<r) =
\begin{cases}
M_{\rm\textsc{cNFW}}(<r),
& r < r_{\mathrm{t}}, \\[8pt]
\begin{aligned}
&M_{\rm\textsc{cNFW}}(r_{\mathrm{t}})\ +\\
&4\pi \rho_{\mathrm{cNFW}}(r_{\mathrm{t}})\,
\dfrac{r_{\mathrm{t}}^{3}}{3-\delta}\, \times
\left[
\left(\dfrac{r}{r_{\mathrm{t}}}\right)^{3-\delta}
-1
\right],
\end{aligned}
& r > r_{\mathrm{t}},
\end{cases}
\end{equation}
where $r_{\mathrm{t}}$ is the radius at which mass is tidally stripped
from the satellite, and $\delta$ is the logarithmic density slope beyond $r_{\mathrm{t}}$. Figure~\ref{fig:example_cNFW} includes a \textsc{coreNFWtides} halo for illustration. 

For the eight dwarf spheroidals, we adopt the results derived by \cite{alvarez2020}, who fitted all parameters simultaneously using the priors $8.5\,{<}\log(M_{200}/M_\odot)\,{<}\,10.5$, $9\,{<}\,c_{200}\,{<}\,24$, $-2\,{<}\log(r_{\rm c}/\rm{kpc})\,{<}\,0.5$, $0.3\,{<}\log(r_{\rm t}/R_{\rm{eff}})\,{<}\,0.5$, $3.5\,{<}\delta\,{<}\,5$, and $0\,{\leq}\,n\,{\leq}\,1$. As discussed in \cite{sarrato2026_dsph}, the best-fitting parameters produce density profiles in broad agreement with other literature estimates within the uncertainties.

\section{Results and discussion}
\label{sec:results_discussion}

\subsection{Dark matter cores and degree of coreness}
\label{sec:cores}
In this section, we examine the distributions of $n$ and $\eta$ as functions of $M_\ast$, which spans the range $4\,{\times}\,10^5\, {\lesssim}\, M_\ast/M_\odot\, {\lesssim}\, 2\,{\times}\,10^{11}$. We work with $\eta$ rather than $r_{\rm c}$ because for some models $r_{\rm c}$ depends by construction on $R_{\rm eff}$, which correlates with $M_\ast$, and may therefore introduce artificial correlations due to various galaxy properties that vary with $M_\ast$, such as a $r_{\rm c}\,{-}\, M_\ast$ trend.

\begin{figure}
    \centering
    \includegraphics[width=1\linewidth]{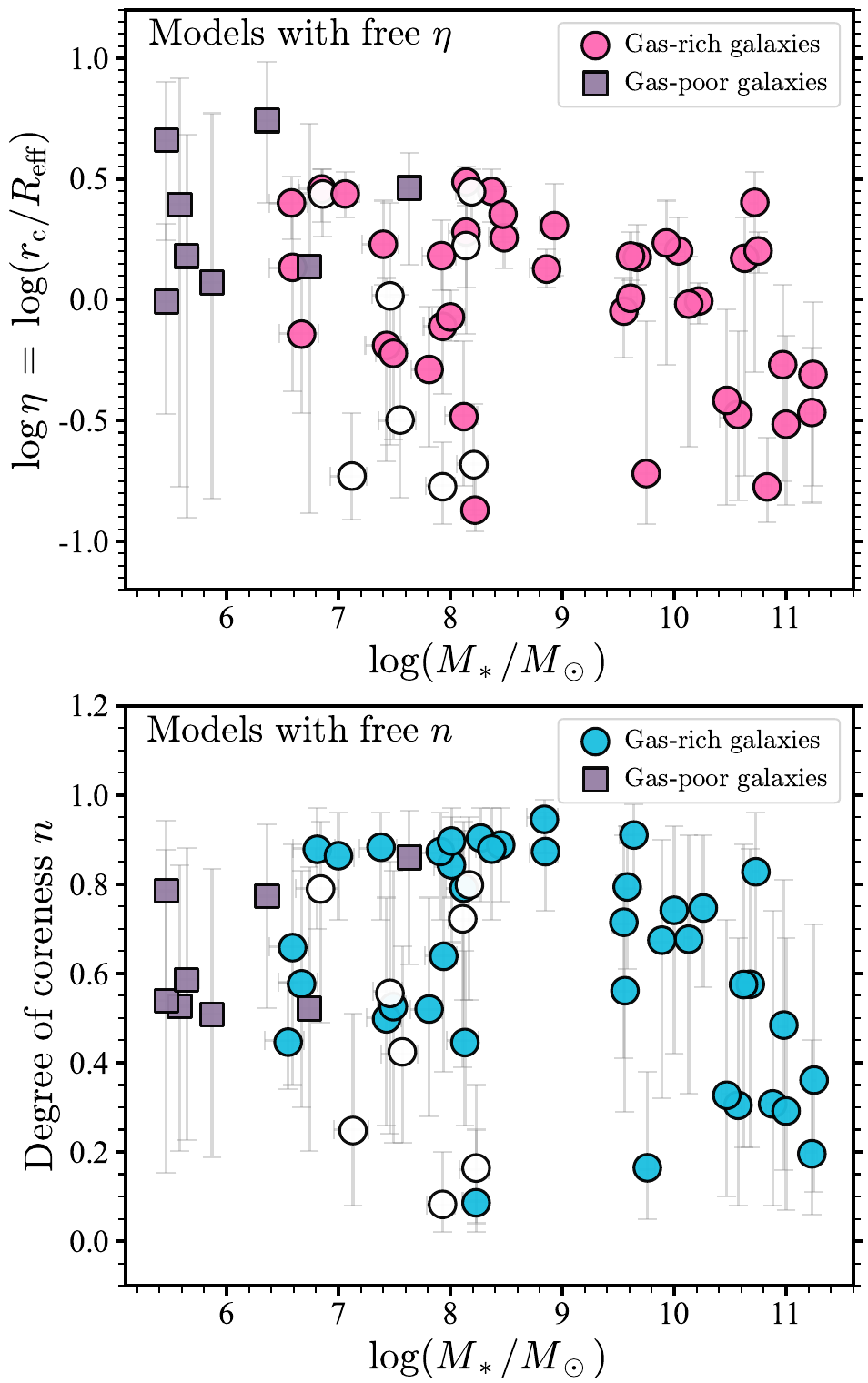}
    \caption{Distribution of $\eta$ and $n$ as functions of $M_\ast$. In both panels, circular markers indicate the median of the best-fitting posterior distributions. White markers correspond to disc galaxies with fewer than two rotation curve measurements within $R_{\rm eff}$. 
    }
    \label{fig:mstar}
\end{figure}

Figure~\ref{fig:mstar} presents the $\eta{-}M_\ast$ and $n{-}M_\ast$ planes. The former uses the mass models introduced in Section~\ref{sec:original_models}, while the latter uses those from Section~\ref{sec:new_models}. We show the median of the posterior distributions, along with the $16\rm{th}{-}84\rm{th}$ percentile range as uncertainties\footnote{
Some of the posterior distributions of $n$ and $\eta$ are skewed towards their prior bounds. This gives rise to the asymmetric uncertainties shown in Figure~\ref{fig:mstar} and also means that some medians are not necessarily the maximum-likelihood values. Still, they capture the main behaviour of $n$ and $\eta$ as a function of $M_\ast$.}. We also examine galaxies with fewer than two rotation-curve measurements within $R_{\rm eff}$, as their inner dark matter densities are more prone to observational uncertainties; however, we do not find these galaxies to be statistically different from more-resolved systems.

Although more massive galaxies tend to be cuspier, both planes in Figure~\ref{fig:mstar} show significant scatter at all $ M_\ast$. We have corroborated that this scatter is not due to poor fits. Instead, it is a consequence of the observed innermost circular velocities, which are rather high in the cuspier galaxies. It is important to stress that the kinematic models of these systems do not show compelling evidence suggesting the inner velocities have been overestimated (see \citealt{iorio,enrico_radialmotions}; \citetalias{paper_galaxyhalo} for details). Therefore, the scatter appears real and is reminiscent of the CDM diversity problem in rotation curves. In the following subsections, we assess whether the inferred cores are energetically feasible, compare the dark matter densities and logarithmic slopes with those of hydrodynamical simulations, and discuss the principal observational and theoretical caveats.

\subsection{Energetics of core formation}
\label{sec:energy}

A relevant question related to our results is whether the core sizes and inner dark matter densities are realistic in the context of feedback-induced core creation. Previous studies have argued that under reasonable halo parametrisations, dark matter cores of realistic sizes can be created by stellar feedback with a few per cent of the SNe energy available over the dwarfs' SFHs \citep{maxwell2015,read2017,burger2021,vogl2025,shinozaki2026}.

We compute the change in gravitational potential energy associated with transforming cuspy NFW haloes into \textsc{coreNFW} profiles of the same mass and concentration, including both the dark matter self-gravity and the interaction between the dark matter and the baryonic potential. Defining $\Delta W$ as the final-minus-initial difference, we can write
\begin{equation}
\label{eq:deltaw}
\Delta W
\equiv
W_{\textsc{coreNFW}}-W_{\rm NFW}
=
\Delta W_{\rm DM}
+
\Delta W_{\rm bar-DM},
\end{equation}
where the first term describes the change in the dark matter self-energy and the second describes the change in the interaction energy between the dark matter and the fixed baryonic potential.

Assuming that the initial and final configurations are in virial equilibrium, the virial theorem implies that the mechanical energy required to form the cored profile is
\begin{equation}
\label{eq:ecore}
E_{\rm core}=\frac{\Delta W}{2}.
\end{equation}

For two spherical profiles enclosing the same mass at $R_{200}$, the standard expressions for the gravitational potential energy (Eqs.~2.17 and 2.24 of \citealt{binney}) yield
\begin{equation}
\label{eq:w_dm}
\begin{split}
\Delta W_{\rm DM}
&=-
\frac{1}{2}
\int_{0}^{R_{200}}
\frac{G\left(M_{\textsc{coreNFW}}^{2}(r)-M_{\mathrm{NFW}}^{2}(r)\right)}{r^{2}}
\,dr~.
\end{split}
\end{equation}
This potential-energy difference is commonly used in calculations of the energy associated with cusp--core transformations (e.g. \citealt{penarrubia2012,maxwell2015,read2017}). For galaxies modelled with the \textsc{coreNFWtides} profile, we evaluate Eq.~\ref{eq:w_dm} using the underlying pre-infall \textsc{coreNFW} profile and its pre-infall $R_{200}$.

The second term in Eq.~\ref{eq:deltaw}, $\Delta W_{\rm bar-DM}$, captures the change in the interaction energy between the dark matter and baryonic potentials. This term is seldom considered, as it is not critical for dwarf galaxies, which are usually heavily dark matter dominated, but it can become relevant for massive galaxies and is important to consider for our sample. The expression for $\Delta W_{\rm bar-DM}$ is
\begin{equation}
\label{eq:w_coupled}
\begin{split}
\Delta W_{\rm bar-DM}
&=
\int \Delta\rho_{\mathrm{DM}}(r)\,\Phi_{\rm bar}(r)\,d^3r \\
&=
2\pi \int_0^{R_{\rm max}} \int_{-z_{\rm max}}^{z_{\rm max}}
\Delta\rho_{\rm DM}(R,z)\,
\Phi_{\rm bar}(R,z)\,
R\,dR\,dz,
\end{split}
\end{equation}
where $\Delta\rho_{\rm DM}(r)\,{=}\,\rho_{\textsc{coreNFW}}(r)\,{-}\,\rho_{\rm NFW}(r)$, and $\Phi_{\rm bar}$ is the gravitational potential provided by the baryons. We note that in this first-order calculation, the baryonic potential is fixed to that inferred from the present-day baryonic mass distribution. Eq.~\ref{eq:w_coupled} therefore accounts for the change in the baryon--dark-matter interaction energy caused by the redistribution of the dark matter, but not for changes in the baryonic distribution itself. A fully self-consistent calculation would require knowledge of the evolving stellar and gas distributions and of the energy exchanged over repeated feedback cycles.

We compute Eq.~\ref{eq:w_coupled}\footnote{Assuming spherical symmetry (as we have for the gas-poor dwarfs), integration by parts and ${d}\Phi_{\rm bar}/{d}r\,{=}\, G M_{\rm bar}(r)/r^2$ reduce Eq.~\ref{eq:w_coupled} to
\begin{equation*}
\Delta W_{\mathrm{bar-DM}}
=
-\int_{0}^{R_{200}}
\frac{G\,M_{\rm bar}(r)\left(M_{\textsc{coreNFW}}(r)-M_{\rm NFW}(r)\right)}{r^2}\,dr\ .
\end{equation*}} by sampling the mass models' posterior distributions. For the gas-rich discs, we use the models from Section~\ref{sec:new_models}, but we find results similar to those in Section~\ref{sec:original_models}. The baryonic mass distributions for the gas-rich systems are taken from \citetalias{paper_galaxyhalo} and include the stellar and \hi discs, as well as a bulge and an H$_2$ disc if present. For the gas-poor systems, we assume $M_{\rm bar}\,{=}\, M _\ast$ and that their stellar component is defined by a spherical Plummer halo with $R_{\rm eff}$ and $M_\ast$ as listed in \cite{read2019}. For both $R_{\rm max}$ and $z_{\rm max}$ we adopt $R_{200}$, but we note that the integration limit has negligible consequences, as the \textsc{coreNFW} profile follows an NFW in the outer regions.

We compare $E_{\rm core}$ with a first-order estimate of the SNe energy available for core formation, $E_{\rm SN}$, given by
\begin{equation}
    E_{\rm SN} =
    \dfrac{M_\ast({<}R_{\rm eff})}
    {(1-R_{\rm f})\,\langle m_\ast\rangle}\,
    e_{\mathrm{SN}}\,
    \xi(m_\ast>8\,M_\odot)\,
    \epsilon_{\mathrm{SN}}~.
\end{equation}
Here, $M_\ast({<}R_{\rm eff})$ is the present-day $M_\ast$ within $R_{\rm eff}$, (arguably the fraction involved in the core-formation process); $R_{\rm f}\,{=}\,0.441$ is the fraction of the initially formed stellar mass returned to the interstellar medium through stellar evolution; $\langle m_\ast\rangle\,{=}\,0.83\, M_\odot$ is the average stellar mass; $e_{\mathrm{SN}}\,{=}\,10^{51}\,\rm erg$ is the average energy released per supernova after accounting for energy lost to neutrinos; $\xi(m_\ast\,{>}\,8\, M_\odot)\,{=}\,0.00978$ is the fraction of stars that become SNe; and $\epsilon_{\mathrm{SN}}$ is the effective net coupling efficiency between the total SNe energy budget and dark matter orbital energy, incorporating both energy losses in the ISM and the transfer mediated by feedback-driven gas motions\footnote{Baryonic particles are subject to the same fluctuations in the gravitational potential and may therefore also gain orbital energy. This contribution is not included in $E_{\rm SN}$ as defined here: $\epsilon_{\mathrm{SN}}$ denotes specifically the net fraction of the SNe energy budget transferred to the dark matter.}. The values of $R_{\rm f}$, $\langle m_\ast\rangle$, and $\xi$ assume a \citet{chabrier2003} initial mass function, with $R_{\rm f}$ taken from \citet{spitoni2017}.

\begin{figure}
    \centering
    \includegraphics[width=1\linewidth]{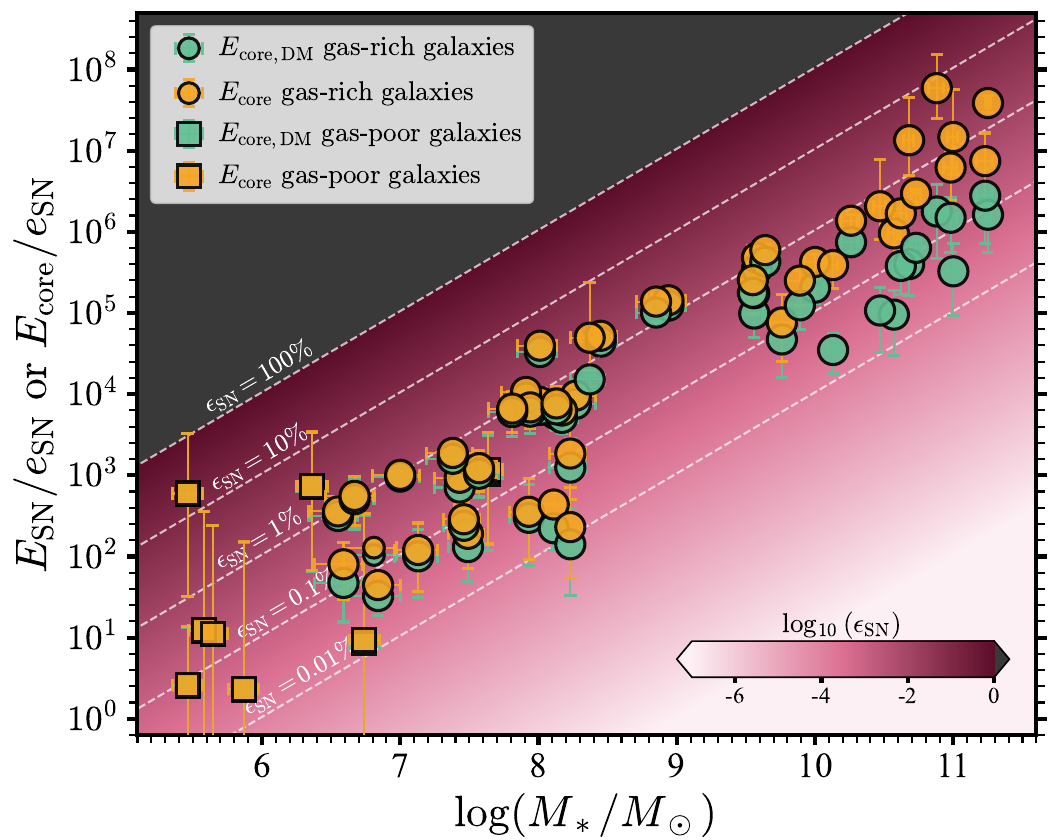}
    \caption{Energy $E_{\rm core}$ needed to transform NFW haloes into the cored profiles of our sample given the dark matter (green) and total potential (yellow). Circles (squares) represent gas-rich (gas-poor) galaxies. The background and white dashed curves show the energy available from SNe ($E_{\rm SN}$) for various coupling efficiencies $\epsilon_{\mathrm{SN}}$. The grey zone denotes a region that requires $\epsilon_{\rm SN}\,{>}\,100\%$. Both $E_{\rm core}$ and $E_{\rm SN}$ are normalised to $e_{\mathrm{SN}}\,{=}\,10^{51}\,\rm erg$.
    }
    \label{fig:energy}
\end{figure}

Figure~\ref{fig:energy} presents the dark-matter-only ($E_{\rm core,DM}\,{\equiv}\,\Delta W_{\rm DM}/2$) and total ($E_{\rm core}\,{\equiv}\,\Delta W/2$) core-formation energies as functions of $M_\ast$. These quantities are compared with $E_{\rm SN}$ evaluated for different values of $\epsilon_{\mathrm{SN}}$. For galaxies with $M_\ast\,{\lesssim}\,10^9\, M_\odot$, $E_{\rm core}\,{\approx}\, E_{\rm core, DM}$, while in more massive systems the baryon--dark-matter interaction makes an important contribution to the required energy. However, even when accounting for the total gravitational potential, the efficiencies required to reproduce the observed dark matter cores and inner density slopes are below $1\%$ for the vast majority of galaxies in our sample. A handful of systems require $\epsilon_{\mathrm{SN}}\,{\approx}\,1{-}5\%$, and only Ursa Minor needs a larger value
($\epsilon_{\mathrm{SN}}\,{\approx}\,20\%$). The dwarf irregular IC~2574 (with kinematics derived in \citetalias{paper_galaxyhalo}), one of the most well-known cases with large dark matter cores \citep{oman2015}, requires only $\epsilon_{\mathrm{SN}}\,{\approx}\,2\%$.

Within the assumptions of our energetic calculation, modest\footnote{
Our estimate includes only core-collapse SNe. The time-integrated number per unit stellar mass of SNe Ia is about $90\%$ lower than that for core-collapse SNe, and their delayed, less clustered explosions are expected to be less effective at driving the rapid changes in gravitational potential needed for core formation \citep{maoz2012,fielding2018}. AGN feedback may noticeably affect galaxy baryon fractions \citep{contini2025}, but its impact on dark matter density profiles remains debated \citep{koudmani2025}. Our energy estimates may therefore be conservative.
} effective SNe coupling efficiencies are sufficient to reproduce the full range of dark matter density profiles spanned by our galaxy sample across six orders of magnitude in $M_\ast$. These efficiencies describe the net transfer from the total SNe energy budget to dark matter orbital energy, encompassing the intermediate coupling through feedback-driven gas motions. Although the efficiency of this transfer depends on the amount, spatial concentration, and timescale of the gas redistribution \citep{pontzen2012}, the low values required across our sample indicate that the available core-collapse SNe energy budget can readily account for the observed dark matter distributions.

\subsection{Comparison with $\Lambda$CDM hydrodynamical simulations}
\label{sec:simulations}

The scatter observed in Figure~\ref{fig:mstar}, reminiscent of the diversity-of-rotation-curves problem, has two contributions: some intrinsic to galaxy evolution and some due to observational uncertainties. By working with a curated sample of galaxies, we aimed to minimise the latter (see also the discussion in Section~\ref{sec:caveats}), but significant scatter remains. 

Recent studies based on hydrodynamical simulations have shown that the creation of dark matter cores depends on the duration and burstiness of galaxies' SFHs, as well as on the fraction of their $M_\ast$ formed before and after the epoch of re-ionisation (e.g. \citealt{onorbe2015,read2019,azartash2024,muni2025,sarrato2026}). Late, extended, and bursty SFHs provide more effective and repeated dark matter heating, resulting in more efficient core formation at fixed $M_\ast$. In this context, the scatter seen in Figure~\ref{fig:mstar} could be due to galaxies with a range of SFHs and cumulative core-formation histories being at different stages of their core creation process. 

To evaluate whether these types of mechanisms can explain our observations and assess the degree to which we are witnessing a manifestation of the diversity-of-rotation-curves problem, we compare the halo properties of our galaxy sample against results from the FIRE-2 \citep{fire2,graus2019,wheeler2019}, NIHAO \citep{nihao_sims}, and EDGE-1 and EDGE-2 (hereafter simply EDGE) simulations \citep{edge,orkney2021,rey2025}, all of which are capable of forming feedback-driven dark matter cores.

For the data, we use the same models analysed in Figure~\ref{fig:energy} and complement them with a model of the Small Magellanic Cloud (SMC) derived by \cite{deleo2024} using the \textsc{coreNFWtides} profile. The SMC is an interesting case, as it is a low-mass system that appears to retain a dark matter cusp in a mass regime where core formation is expected to be maximal. For each galaxy, we extract the logarithmic slope of the dark matter density profile ($\alpha_{\rm DM}\,{=}\,d\ln\rho_{\rm DM}/d\ln r$) and dark matter volume density $\rho_{\rm DM}$, both evaluated at the same $r_{\rm scale}$ values. For $r_{\rm scale}$, we use $300\,\rm{pc}$ and $0.015\,R_{200}$. These inner scales are chosen to probe different regions where feedback-driven halo modification is expected to be strongest. For completeness, in Appendix~\ref{app:large_scale} we also examine the larger scales of $5\,\rm{kpc}$ and $0.060\, R_{200}$, which generally probe the transition towards a less strongly modified halo. As discussed in Appendix~\ref{app:large_scale}, such diagnostics are also helpful to identify mismatches in the galaxy-halo mapping.

For the simulations, our approach is as follows. For EDGE, we use the dark matter density profiles taken from \cite{muni2025}. For NIHAO, the main isolated halo from each zoom-in simulation was selected if it contained at least 100 star particles, resulting in 93 galaxies from 93 zoom-in runs. From FIRE-2, 109 galaxies were selected by keeping the ten haloes with the largest particle counts in each FIRE-2 zoom-in volume, as long as they were composed of at least 99\% high-resolution particles and contained a minimum of a hundred star particles. For all simulations, $\alpha_{\rm DM}$ is determined by performing a linear fit to the binned $\log\rho_{\rm DM}$--$\log r$ relation over $0.9\,r_{\rm scale}<r<1.1\,r_{\rm scale}$, while $\rho_{\rm DM}(r_{\rm scale})$ is measured from the density bin spanning the same radial interval. We also obtain the value of the $n$ parameter from \textsc{coreNFW} fits to the density profiles, and further discuss it in Appendix~\ref{app:nparameter}.

Lastly, we compare data and simulations against a family of reference CDM haloes. We generate a mock halo population following the $z\,{=}\,0$ halo mass function of \cite{tinker2008} and assign stellar masses using the $M_\ast{-}M_{\rm vir}$ relation of \cite{moster2010}, including a scatter of 0.15 dex\footnote{
We adopt this standard SHMR for simplicity. However, evidence suggests that blue late-type and red early-type galaxies follow different SHMRs (e.g. \citealt{aldo2015,posti2021}). At high $M_\ast$, where early-type galaxies dominate the galaxy population, their SHMR exhibits a high-mass bend similar to that of the standard relation of \cite{moster2010}. In contrast, at least some massive disc galaxies appear to follow a linear relation between $\log(M_\ast)$ and $\log(M_{200})$, without a high-mass bend (e.g. \citealt{posti_galaxyhalo,enrico_massmodels_ss}).}, and accounting for the $M_{\rm vir}\,{\rightarrow}\, M_{200}$ transformation. The haloes are assumed to be NFW, with concentrations drawn from the $c_{\rm 200}{-}M_{\rm 200}$ relation of \cite{diemer2019}, with a scatter of 0.16 dex. 
Given that we subsequently compare dark matter densities and slopes at fractional scales of $R_{200}$ and physical kiloparsec scales, it is interesting to note a property arising from the self-similarity
of dark matter haloes. At a given redshift, the density and logarithmic slope of an NFW profile at $r\,{=}\,r_{\rm scale}$ can be written as
\begin{equation}
\rho_{\rm NFW}(r_{\rm scale})=
\frac{200}{3}\,\rho_{\rm crit}\,
\frac{c_{200}^{3}}{g(c_{200})}
\frac{1}{x\,(1+x)^2},\ \rm{and}
\end{equation}
\begin{equation}
\alpha_{\rm NFW}(r_{\rm scale})
\equiv
\frac{{d}\ln\rho_{\rm NFW}}{{d}\ln r}
=
-1-\frac{2\,x}{1+x},
\end{equation}
with $x\,{\equiv}\,r_{\rm scale}/r_{\rm s}\,{=}\,r_{\rm scale}\,c_{200}/R_{200}$ and
$g(c_{200})\,{=}\,\ln(1\,{+}\,c_{200})-[c_{200}/(1+c_{200})]$.

\begin{figure*}
    \centering
    \includegraphics[width=17cm]{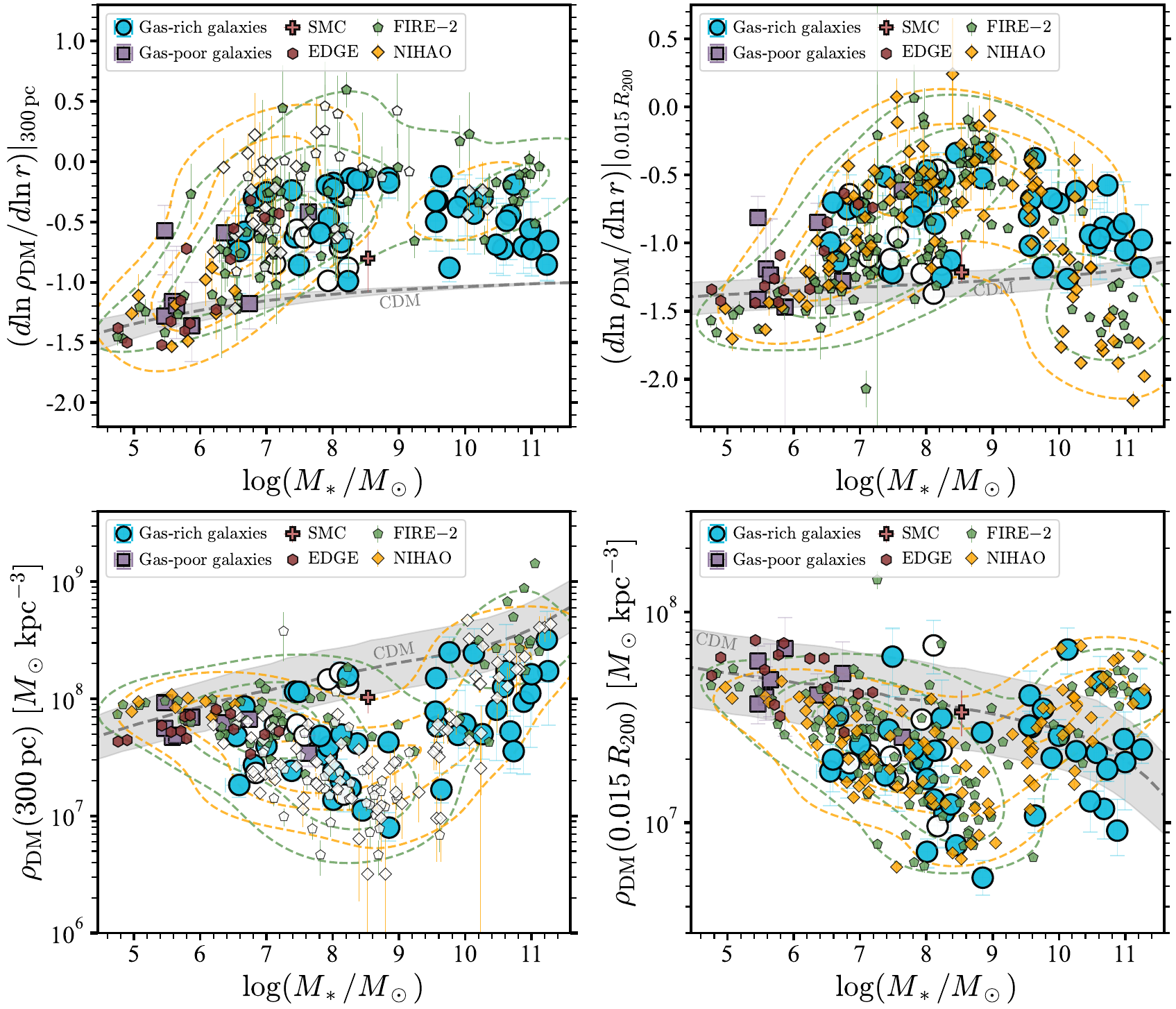}
    \caption{
    Inner dark matter logarithmic slope (top) and density (bottom) measured at 300 pc (left) and $0.015\, R_{200}$ (right) for our observational sample (symbols as in Figure~\ref{fig:mstar}) and comparison simulations. 
    The coloured dashed contours enclose the highest-density regions containing 15\%, 50\%, and 85\% of the KDE probability for FIRE-2 (green) and NIHAO (yellow).
    The grey dashed curve and its shaded region represent a comparison CDM population and its 16th–84th percentile range. }
    \label{fig:simulations_inner}
\end{figure*}

For a given scale $r_{\rm scale}\,{=}\,f\,R_{200}$, with $f$ an arbitrary fraction, one has $x\,{=}\,f\,c_{200}$. Thus, at fixed
$f$, neither $\rho_{\rm NFW}$ nor $\alpha_{\rm NFW}$ depends explicitly
on $M_{200}$; halo mass enters only indirectly through the relatively weak concentration--mass relation, which has $c_{200}\,{\propto}\, M_{200}^{-0.1}$ over the relevant mass range.
By contrast, since $R_{200}\,{\propto}\, M_{200}^{1/3}$, at a fixed physical radius
$x\,{=}\,c_{200}\,r/R_{200}\,{\propto}\, M_{200}^{-0.43}$. A fixed physical radius consequently probes systematically different regions of the NFW profile in haloes of different mass, allowing more pronounced trends of both $\rho_{\rm NFW}$ and $\alpha_{\rm NFW}$ with $M_{200}$. Measurements at
fixed fractions of $R_{200}$ therefore more directly probe departures from the self-similar structure of the haloes, including their response to baryonic processes. Measurements at fixed physical radii additionally depend on the mass-dependent size of the halo and, when examined as a function of $M_\ast$, on the galaxy--halo mass mapping (i.e. the SHMR). With all this in mind, we now compare the distributions of the logarithmic slopes and inner dark matter densities as functions of $M_\ast$ for observations and simulations.

In Figure~\ref{fig:simulations_inner}, we show the inner logarithmic slopes (top panels) and dark matter densities (bottom panels) at 300 pc (left panels) and $0.015\, R_{200}$ (right panels). Figures~\ref{fig:simulations_inner_m200} and \ref{fig:simulations_inner_fstar} present the same quantities as functions of $M_{200}$ and $M_\ast/M_{200}$, respectively. At both radii, the galaxies trace a U-shaped dependence of density on $M_\ast$, together with an approximately inverted U-shaped trend in slope, and depart visibly from the unmodified NFW reference over $6.5\,{\lesssim}\,\log(M_\ast/M_\odot)\,{\lesssim}\,10.5$. 
We note that at the lowest $M_\ast$, most gas-poor satellites lie relatively close to the unmodified NFW reference. At first sight, this may appear inconsistent with their $n$ and $\eta$ values shown in Figure~\ref{fig:mstar}. However, for these systems, $n$ and $\eta$ were fitted simultaneously, whereas Figure~\ref{fig:mstar} shows the separate marginal posterior distributions of the two parameters, which have broad uncertainties and whose medians need not correspond to the same posterior realisation. The quantities in Figure~\ref{fig:simulations_inner} are instead derived from the full posterior profiles at the specified radii and account for model degeneracies. Moreover, $r_{\rm c}$ sets the scale over which the \textsc{coreNFW} profile transitions back towards NFW, while $0<n<1$ describes only a partial core transformation. The local slopes and densities at 300 pc or $0.015\, R_{200}$ can therefore remain close to NFW, depending on where these radii lie relative to $r_{\rm c}$.

Remarkably, the hydrodynamical simulations reproduce not only the systematic trends and the departure from NFW (attributed to stellar feedback) at both 300 pc and $0.015\, R_{200}$, but also the broad distribution of observed values (perhaps even broader, but consistent within the uncertainties). Some simulated galaxies have resolution limits larger than 300 pc, as indicated by the white symbols, making their measurements at this radius, particularly their logarithmic slopes, more susceptible to numerical effects (see \citealt{sarrato2026_dsph}). The agreement at 300 pc should therefore be interpreted with some caution, although the resolved systems remain fully compatible with the observations.

While the low- and intermediate-mass regimes show very good agreement, we identify some systematic differences at the high-mass end, particularly in the slopes for galaxies with $\log(M_\ast/M_\odot)\,{\gtrsim}\,10.5$. At 300 pc, the simulations have slopes less steep than the data, although they remain consistent within the uncertainties. The differences are more significant at $0.015 \, R_{200}$, where simulations produce massive galaxies with slopes steeper than the NFW CDM reference and the data, accompanied by a modest enhancement in dark matter density. 

We surmise that this may be a signature of baryon-driven contraction. \cite{tollet2016} showed that massive NIHAO galaxies can accumulate substantial stellar mass in their central regions, deepening the total gravitational potential and causing their dark matter haloes to recontract. This process can erase previously formed cores and produce steep dark matter profiles. These steep slopes are not apparent at 300 pc, which may indicate that feedback-driven flattening persists at the smallest radii while contraction dominates farther out. However, the limited numerical resolution at 300 pc prevents a firm conclusion.

If comparable contraction occurs in real galaxies, it should be encoded in their present-day dark matter distributions. Recovering this signal from rotation curves is nevertheless challenging because massive disc galaxies are baryon-dominated at these radii, making the inferred dark matter profile particularly sensitive to the adopted mass-to-light ratios and bulge and disc decompositions, as further discussed in Section~\ref{sec:caveats}. Moreover, although our \textsc{coreNFW} models can describe cored or NFW-like haloes, they contain no independent freedom to represent an enhancement of the enclosed inner dark matter mass relative to NFW at fixed $M_{200}$ and $c_{200}$. A contracted halo could therefore be absorbed partly through shifts in the inferred concentration, halo mass, or stellar mass-to-light ratio. Approximate contraction prescriptions and numerical methods exist (e.g. \citealt{blumenthal1986,sellwood2005}), and rotation curve fits that include the response of the dark matter to the observed baryonic distribution can yield slopes between $-1.5$ and $-2$ at $r_{\rm scale}\,{\approx}\,0.015\, R_{200}$ in massive disc galaxies \citep{li_adiabatic}. This shows that slopes as steep as those found in the simulations may be compatible with observations, although it does not imply that contraction should simply be applied as a correction to our inferred profiles. AGN feedback may further complicate the comparison by opposing contraction through the redistribution of central gas and dark matter, particularly since it is not included in the simulations considered here. The apparent discrepancy at $\log(M_\ast/M_\odot)\,{\gtrsim}\,10.5$ should therefore be interpreted cautiously. In future work, we plan to test it using halo models that permit both expansion and contraction while simultaneously exploring uncertainties in the stellar mass distribution.

Taken together, the panels of Figure~\ref{fig:simulations_inner} show that the simulations largely reproduce both the systematic mass dependence and the observed galaxy-to-galaxy variation in the inner dark matter slopes and densities. This substantially alleviates the diversity-of-rotation-curves problem within our curated sample: we find no systematic population of galaxies with inner dark matter densities below (or generally significantly more scattered than) those produced by the hydrodynamical simulations. 

As discussed in Appendix~\ref{app:large_scale}, the comparison between data and simulations at larger scales reveals some differences in the SHMR between data and simulations and even among different simulations (somewhat analogous to reported discrepancies in the $M_\ast\,{-}\,R_{\rm e}$ relation). Yet, as discussed there, such dissimilarities are more directly related to the global galaxy-halo mapping rather than to a systematic disagreement in the inner dark matter densities at fixed $M_{200}$. Despite the remaining uncertainty in the SHMR and its intrinsic scatter (see \citealt{sales_review_dwarfs}), our central result is the agreement between the dark matter slopes and densities inferred from observations and those measured in hydrodynamical simulations. This agreement is found both at homologous halo radii and at fixed physical radii.

Together with the energetic arguments presented in Section~\ref{sec:energy}, our results substantially alleviate the diversity-of-rotation-curves problem. Although the observed inner dark matter densities span a wide range, this diversity is broadly consistent with hydrodynamical $\Lambda$CDM simulations in which core formation and the resulting galaxy-to-galaxy scatter depend on the duration and burstiness of the SFHs and the fraction of stellar mass formed after re-ionisation. We therefore find no evidence of a systematic mismatch in the inner dark matter structure between observed and simulated galaxies. Looking ahead, it will be crucial to obtain homogeneously reconstructed SFHs (e.g. \citealt{mcquinn2026}) and improved halo-mass constraints to determine whether the observed galaxy-to-galaxy scatter follows the correlations with SFH properties predicted by the simulations.

\subsection{The $(\eta_{\rm bar},\eta_{\rm rot})$ plane}
\label{sec:comparison}

As mentioned above, several studies have investigated the diversity-of-rotation-curves problem. A particularly relevant work is that by \citet[][hereafter \citetalias{santos2020}]{santos2020}, who analysed 160 late-type galaxies spanning approximately five orders of magnitude in $M_\ast$. In contrast to the results presented in the previous section, those authors reported significant tension between observations and hydrodynamical simulations.

 \begin{figure*}[h]
\sidecaption
  \includegraphics[width=12cm]{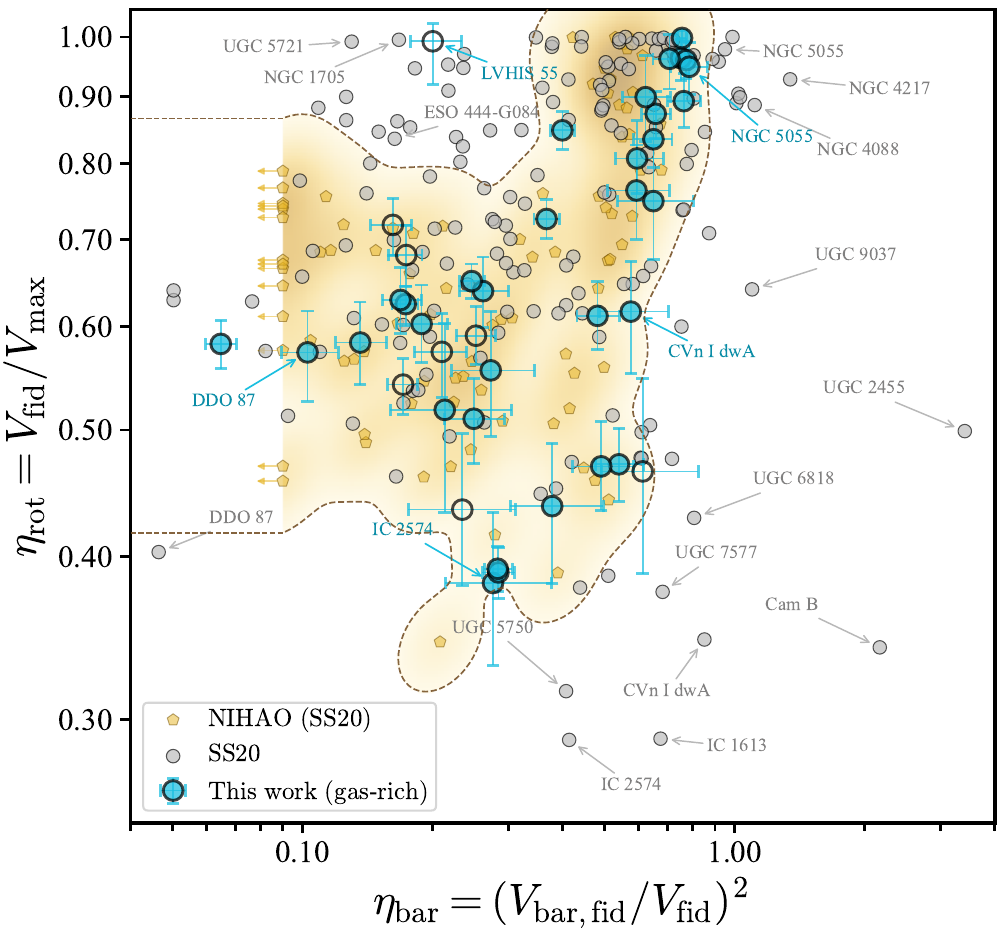}
     \caption{
     Distribution of galaxies in the $(\eta_{\rm bar},\eta_{\rm rot})$ plane. Grey and yellow markers show the observational and NIHAO samples of \citetalias{santos2020}, respectively, while cyan markers show the gas-rich galaxies analysed in this work (open symbols as in Figure~\ref{fig:mstar}). The yellow background and dashed contour represent the density of NIHAO galaxies in this parameter space. Note that \citetalias{santos2020} truncated NIHAO galaxies with $\eta_{\rm bar}\,{<}\,0.09$ to $\eta_{\rm bar}\,{=}\,0.09$, so we display those measurements as upper limits and keep the contour open.
        Selected galaxies discussed in the text are labelled.
     }
     \label{fig:santos}
\end{figure*}

\citetalias{santos2020} assessed the agreement between observations and simulations in the $(\eta_{\rm bar},\eta_{\rm rot})$ plane, where
\begin{equation}
\eta_{\rm bar}
=
\left(\frac{V_{\rm bar,fid}}{V_{\rm fid}}\right)^2,
\qquad
\eta_{\rm rot}
=
\frac{V_{\rm fid}}{V_{\rm max}}.
\end{equation}
Here, $V_{\rm max}$ is the maximum circular speed in the rotation curve (after correcting for asymmetric drift), while $V_{\rm fid}\,{=}\, V_{\rm circ}(r_{\rm fid})$ is the circular speed measured at the fiducial radius
\begin{equation}
r_{\rm fid}
= 2\left(\frac{V_{\rm max}}{70\,{\rm km\,s^{-1}}}\right)\,{\rm kpc}~,
\end{equation}
and $V_{\rm bar, fid}\,{=}\, V_{\rm bar}(r_{\rm fid})$ is the total baryonic contribution to the circular speed evaluated at the same radius. Therefore, $\eta_{\rm rot}$ quantifies the inner shape of the rotation curve: low values correspond to slowly rising rotation curves, whereas values approaching unity indicate rotation curves that rise rapidly and have already reached close to their maximum speed at $r_{\rm fid}$. Meanwhile, $\eta_{\rm bar}$ measures the fractional contribution of baryons to the gravitational acceleration at $r_{\rm fid}$. While $\eta_{\rm bar}$ and $\eta_{\rm rot}$ are useful diagnostics, they are not statistically independent: both depend on $V_{\rm fid}$, while $V_{\rm max}$ determines $r_{\rm fid}$ and hence the radius at which both $V_{\rm fid}$ and $V_{\rm bar, fid}$ are evaluated. Uncertainties or systematic biases in $V_{\rm max}$ and $V_{\rm fid}$ therefore propagate into both coordinates, producing correlated errors \citep{roeper2023}.

\citetalias{santos2020} identified regions in the $(\eta_{\rm bar},\eta_{\rm rot})$ plane that are populated by real galaxies but not by simulations, specifically APOSTLE \citep{apostle_sawala,fattahi_apostle}, EAGLE \citep{eagle,eagle_crain}, and NIHAO. NIHAO came closest to reproducing the data but still failed to reproduce the full observed distribution of rotation-curve shapes. Similar results are found by \cite{jackson2025} using the NEWHORIZON simulations \citep{newhorizon}, and by \cite{cruz2026} with the Marvel/Marvelous simulations \citep{bellovary2019,munshi2021,christensen2024}. In particular, simulations consistently miss galaxies matching those from \citetalias{santos2020} with slowly rising rotation curves and high baryon fractions, as well as steeply rising rotation curves and low baryon fractions. At fixed $M_\ast$, this would roughly correspond to missing simulated systems with lower and higher inner $\rho_{\rm DM}$ than real galaxies, respectively. Clearly, this contrasts with our findings in Figure~\ref{fig:simulations_inner}.

To investigate this discrepancy, in Figure~\ref{fig:santos} we show the $(\eta_{\rm bar},\eta_{\rm rot})$ plane. We include the observational sample from \citetalias{santos2020}, together with their results for NIHAO; note that NIHAO galaxies with $\eta_{\rm bar}<0.09$ were assigned a value of $\eta_{\rm bar}\,{=}\,0.09$ by \citetalias{santos2020}, so we consider them lower limits. Moreover, we include in the figure our own late-type galaxies; following \citetalias{santos2020}, we only include galaxies whose rotation curves extend to $r_{\rm last}\,{\geq}\,2r_{\rm fid}$, where $r_{\rm last}$ is the outermost measured radius.

Clearly, our sample avoids the same regions of the parameter space as NIHAO, whose distribution is highlighted with a 2D density background. The only possible exception is LVHIS~55, whose rotation curve is less robustly constrained because of its relatively sparse spatial sampling (see Section~\ref{sec:cores} and \citetalias{paper_galaxyhalo} for further discussion of its kinematics). Moreover, the bulk of our sample overlaps with the main distribution of observed galaxies from \citetalias{santos2020}. The difference in conclusions regarding the overlap with NIHAO stems primarily from galaxies occupying the most extreme regions of the diagram, which are absent from our curated sample. This suggests that the reported tension may be particularly sensitive to including galaxies with uncertain kinematics or baryonic mass distributions. In fact, we note that several galaxies in the \citetalias{santos2020} compilation have $\eta_{\rm bar}\,{>}\,1$, formally implying that the baryonic contribution to the radial acceleration exceeds the total inferred from the kinematics at $r_{\rm fid}$, suggesting that observational or modelling uncertainties affect at least some of the measurements.

Given this, we first examine the 18 galaxies common to the two samples. For most of these systems, the derived values are qualitatively consistent, but notable differences are found for a handful of them, of which DDO~87, IC~2574, CVn~I~dwA, and NGC~5055 are particularly interesting and are highlighted in Figure~\ref{fig:santos}. For the three dwarf galaxies (DDO~87, IC~2574, CVn~I~dwA), the differences largely stem from the adopted kinematic analyses. \citetalias{santos2020} used rotation curves from the pioneering work on low-mass galaxies by \citet{oh2015}, derived by fitting two-dimensional velocity fields. By contrast, our analysis uses the results of \citet{iorio} and \citetalias{paper_galaxyhalo}, who modelled the full three-dimensional data cubes. As shown by \citet{iorio}, modelling the full data cube produces significant differences in the recovered rotation curves of several galaxies analysed by \cite{oh2015}. Additionally, our mass models account for the flaring of the \hi disc, which can substantially affect $V_{\rm bar}$ in low-mass, gas-rich dwarfs (\citetalias{paper_galaxyhalo}).
For NGC~5055, the difference arises mainly from the adopted baryonic mass distribution. \citetalias{santos2020} used the one-dimensional, non-parametric bulge-disc decomposition from \cite{sparc}, whereas we use the two-dimensional photometric decomposition of \citet{salo2015}, who finds no bulge in this galaxy. In all four cases, the revised kinematic or baryonic modelling moves the galaxies towards less extreme locations and into regions populated by NIHAO.

We next perform a non-exhaustive inspection of several extreme galaxies in the \citetalias{santos2020} sample (not included in ours) that fall in exclusion zones avoided by simulations. These systems are also labelled in Figure~\ref{fig:santos}. While most galaxies show some degree of irregularity \citep{richter1994}, each extreme galaxy presents at least one factor that significantly complicates deriving or interpreting their kinematics.

From bottom right to top left:
For IC~1613, the large \hi holes and disturbed central gas distribution make interpreting its inner gas kinematics particularly difficult \citep{readAD,taibi2024}. UGC~5750 has a highly disrupted velocity field with an unreliable rotation curve \citep{thijs1993,deblok2002}. In Cam~B, the rotation speed is comparable to the gas velocity dispersion, requiring an exceptionally large asymmetric-drift correction; moreover, its rotation and velocity dispersion were not modelled jointly, and only an approximate correction for beam smearing was applied \citep{begum2003}. UGC~7577 was not corrected for asymmetric drift despite this correction being non-negligible \citep{swaters09}. UGC~6818 has an asymmetric position-velocity diagram, and it is likely interacting with a faint companion \citep{marc2001a}. UGC~2455 has misaligned optical and kinematic position angles \citep{garrido2002}, and an \hi warp \citep{swaters02}. UGC~9037 has twisted velocity fields and significant radial motions likely linked to a bar \citep{hallenbeck2014}. NGC~4088 has significant non-circular motions and a poor kinematic fit \citep{marc2001a}. NGC~4217 has an inner rotation curve that depends strongly on the adopted and somewhat uncertain inclination \citep{marc2001a}, and so does the deprojection of its stellar mass profile. ESO~444-G084 has an irregular velocity field and significant radial motions \citep{namuba2025}. NGC~1705 is a starburst dwarf with a warped and asymmetric gaseous disc \citep{elson2013}. Finally, UGC~5721 exhibits strong lopsidedness and a highly asymmetric position-velocity diagram \citep{swaters02}.

Although this inspection is not exhaustive, it reveals that the most extreme systems in the parameter space avoided by NIHAO have substantial observational uncertainties. Similar arguments were proposed by \cite{oman2015}, \cite{readAD}, \cite{oman_noncircularmotions}, and \cite{roeper2023}. 
We caution, however, that the exclusion zones considered here are specific to NIHAO and related simulations in which feedback efficiently forms dark matter cores. Simulations that systematically retain cuspier systems can more readily produce rapidly rising, baryon-poor rotation curves, which NIHAO underproduces (\citetalias{santos2020}); because our sample contains no clear examples of this population, our analysis does not directly test this aspect of the diversity problem.

Our results show that the tension previously identified in the $(\eta_{\rm bar},\eta_{\rm rot})$ plane therefore appears to depend strongly on sample selection and on the modelling adopted for a relatively small number of galaxies. When we restrict the comparison to our sample, in which such uncertainties are limited, the regions avoided by NIHAO remain unpopulated. This comparison reinforces our conclusion that there is no systematic mismatch between the observed inner dark matter distributions and those produced by current hydrodynamical simulations, and that the diversity-of-rotation-curves problem is substantially alleviated for galaxies with robustly constrained kinematics and baryonic mass models.

\subsection{Caveats}
\label{sec:caveats}

\subsubsection{Observational uncertainties}

Our galaxy sample was built to minimise the impact of data limitations such as low resolution, uncertain inclinations and distances, non-circular motions, or poor constraints on the baryonic distribution. As demonstrated in Section~\ref{sec:comparison}, this is key for the study of the inner dark matter content of galaxies. However, even in our curated sample, some level of observational error persists.

First, uncertainties remain in the kinematic modelling. There is no unique prescription for estimating uncertainties in the inferred circular speeds (e.g. \citealt{swatersPhD,deblok08,barolo}). For well-resolved nearby galaxies, these uncertainties are often of the order of $5\,{\rm km\,s^{-1}}$ and are intended to capture limitations in the kinematic models, as well as asymmetries between the approaching and receding sides of the galaxies. Finite spatial resolution and correlations between adjacent rotation-curve points may also introduce uncertainties in the recovered inner circular speeds. The kinematic analyses used in this study avoid strong oversampling of the beam (see \citealt{iorio,enrico_radialmotions,paper_galaxyhalo} for details), thereby limiting the impact of the correlation between rotation curve measurements in the mass models \citep{posti2022}. Nevertheless, these uncertainties may render cored and cuspy mass models statistically consistent for some individual galaxies and may contribute to the observed scatter in their inferred inner dark matter distributions \citep{chase2026}. Future observations with the SKA \citep{ska}, complemented by high-resolution stellar or molecular gas kinematics, should further improve the recovery of the innermost rotation curves. 

Second, despite our use of NIR photometry and SED-informed priors, uncertainty in the stellar mass-to-light ratios remains and can be critical for massive galaxies. This is exemplified in Appendix~\ref{app:massive}, where we show how different plausible mass-to-light ratios may weaken or strengthen the evidence for cores in massive discs. Refinements on mass-to-light ratios, bulge-disc decompositions, and high-resolution stellar kinematics will be crucial to judging the apparent discrepancy in the dark matter slopes shown in the upper-right panel of Figure~\ref{fig:simulations_inner}.

And third, there are inherent limitations in using analytical expressions to model the dark matter density profiles. Empirically, we find the \textsc{coreNFW} profile to fit the dynamics of galaxies across six orders of magnitude in $M_\ast$ with modest SNe efficiencies. Yet this profile accounts only for deviations from NFW driven by stellar feedback and has no independent degree of freedom to capture contraction, which can be important at high masses, as discussed in Section~\ref{sec:simulations}. Therefore, it will be important to incorporate such an effect into the \textsc{coreNFW} halo (e.g. \citealt{coreEinasto}) and evaluate its impact in the future.

\subsubsection{Dynamical disequilibrium}

Feedback-driven star-formation cycles can affect rotation curves in two conceptually distinct ways. First, repeated fluctuations in the central gravitational potential can produce genuine, time-dependent changes in the dark matter distribution, such that the growth of a core need not be monotonic \citep{readAD,tollet2016}. Additional fluctuations may be driven by AGN activity, self-gravitating gas clumps, or mergers (e.g. \citealt{ogiya2014,laporte2015,orkney2021,jahn2023,rey2024,dado2026}). Second, the gas itself may depart from dynamical equilibrium during these events, causing the measured rotation curve to differ from the true circular-speed curve associated with the instantaneous gravitational potential. The sign and magnitude of this bias depend on the galaxy's dynamical state: non-equilibrium motions can lead to either overestimates or underestimates of the circular speed, although recent simulations find a tendency towards underestimation in the inner regions \citep{jahn2023,roeper2023,dado2026}. Mass models based on such kinematics may consequently infer incorrect inner dark matter densities, including apparently cored profiles for intrinsically cuspy haloes.

Several studies have therefore argued that disequilibrium and non-circular motions could account for part of the observed diversity of rotation curves and alleviate some apparently extreme cusp--core discrepancies \citep{oman2015,oman_noncircularmotions,santos2020,downing2023,roeper2023,jahn2023,sands2026,dado2026}. In FIRE-3, \cite{sands2026} find that rotation curves of galaxies with well-ordered gaseous discs generally recover the true circular speed to within approximately $10\%$, but non-equilibrium systems can show substantially larger deviations. Moreover, some cosmological simulations produce low-mass gas discs that are more turbulent and less rotationally supported than their observed counterparts \citep{benavides2025,benavides2026}. Such simulations may therefore overestimate the frequency or amplitude of disequilibrium effects in observational samples of regularly rotating galaxies.

Our gas-rich sample was explicitly curated to exclude galaxies with strong non-circular motions, disturbed kinematics, or other clear signs of disequilibrium, substantially reducing the likelihood of large kinematic biases. Furthermore, the dispersion in the dark matter profiles measured directly in the simulations, where it reflects genuine differences in halo structure rather than errors in recovering the gravitational potential, is comparable to that inferred for our galaxies. This is consistent with much of the observed scatter being intrinsic. We therefore do not require dynamical disequilibrium to explain the diversity of inner dark matter content in our sample, while recognising that it may still affect individual galaxies or contribute to the measured scatter.

\subsubsection{Simulated galaxies}

The agreement between the dark matter densities inferred for our galaxies and those measured in NIHAO, FIRE-2, and EDGE is encouraging, but its interpretation is limited by both sample matching and numerical modelling. First, we use the FIRE-2 and NIHAO subsamples analysed by \cite{sarrato2026}, selected for their numerical resolution but not explicitly matched to the observed sample in morphology, gas content, kinematics, or environment (although all are central galaxies). Moreover, these zoom-in suites are not volume-complete and were not constructed using the same selection function as the observations. A more controlled comparison would apply comparable selection criteria to observations and simulations and forward-model mock kinematic data accounting for observational effects. These mocks would also help quantify biases arising from non-circular motions, observational resolution, and halo-profile inference.

Second, state-of-the-art simulations remain subject to finite resolution and uncertain prescriptions for unresolved physical processes, both of which can affect the formation of dark matter cores (e.g. \citealt{smith2018,ludlow2023,azartash2024,zhang2024,sarrato2026}). For instance, FIRE-2 forms stars with $100\%$ efficiency per local free-fall time in dense, self-gravitating gas \citep{fire2}. Although the galaxy-wide star-formation efficiency is subsequently regulated by stellar feedback, the spatial and temporal clustering of star formation (and hence the fluctuations in the gravitational potential responsible for core formation) remain sensitive to the adopted star-formation and feedback prescriptions. NIHAO instead employs a blast-wave feedback model in which radiative cooling is artificially suppressed temporarily to limit numerical overcooling of the injected SN energy \citep{dutton2016}. The runs considered here also exclude dynamically coupled cosmic rays or black-hole feedback, which may alter star formation and inner-halo structure at high masses and potentially in at least some dwarf galaxies \citep{martizzi2013,koudmani2025}. More generally, these suites do not treat the cold neutral ISM uniformly. Its density and phase structure can affect the spatial and temporal clustering of star formation, and hence the fluctuations in the gravitational potential that drive dark matter heating. Its treatment also determines how faithfully mock \hi observations can trace the underlying gravitational potential. The high resolution of EDGE allows individual SNe to be modelled as discrete events and their impact on the surrounding interstellar medium to be captured without artificially delaying radiative cooling. The updated EDGE-2 implementation additionally follows the non-equilibrium thermochemistry of \hi, ionised hydrogen, and H$_2$ coupled to radiative transfer \citep{rey2025}. However, its sample is smaller and restricted to a lower-mass regime. Moreover, radiative feedback remains less accurately captured than SN feedback, while the possible effects of rare stellar-evolution channels, black-hole feedback, and cosmic rays have not yet been fully explored \citep{rey2025}. These differences in numerical methods and physical prescriptions may contribute not only to differences in the predicted halo response but also to the offsets in the SHMR among the simulations discussed in Appendix~\ref{app:large_scale}.

Consequently, the agreement found in Sect.~\ref{sec:simulations} should not be interpreted as demonstrating that the relevant baryonic physics is uniquely or completely modelled. Compensating systematic effects in either the simulations or the observational inference could contribute to the agreement. Nevertheless, the fact that compatible halo structures emerge across several independent simulation suites employing different numerical methods and feedback prescriptions is nontrivial. Our conclusion should therefore be understood as conditional on the samples, radial scales, and modelling assumptions considered here. Within these limits, we find no systematic mismatch between the inferred and simulated inner dark matter slopes and densities. 

\looseness=-1
Establishing whether this agreement constitutes a robust physical prediction will require improved modelling of feedback and the cold neutral ISM, as well as more closely forward-modelled comparisons between simulated and observed galaxies. It will also be important to determine whether simulated galaxies in which feedback has formed cores would satisfy the same selection criteria as our observed sample when mock-observed, particularly the requirement of regularly rotating \hi discs. Core-forming episodes may precede the present-day observations and allow the gas disc to settle, but if they leave long-lived kinematic disturbances, sample selection could affect the comparison. In this respect, applying the same selection criteria and full observational analysis to mock \hi data cubes from high-resolution zoom simulations such as EDGE-2 and from well-resolved galaxies in representative-volume simulations with an explicitly modelled cold multiphase ISM, such as COLIBRE \citep{colibre}, would test both whether the simulated systems would enter our curated sample and whether the inferred rotation curves recover the underlying dark matter density profiles.

\section{Summary and conclusions}
\label{sec:conclusions}

In this study, we have characterised the dark matter density profiles of a selected sample of 48 nearby late-type disc galaxies and eight Milky Way dwarf satellites, spanning six orders of magnitude in $M_\ast$. Our main findings can be summarised as follows.

\begin{itemize}
\item We find substantial galaxy-to-galaxy scatter in the inferred dark matter core sizes and degrees of coreness at fixed $M_\ast$ (Figure~\ref{fig:mstar}). This is closely linked to the cusp-core and diversity-of-rotation-curves problems, which have remained central challenges for CDM.

\item The inferred, often cored, dark matter density profiles are energetically compatible with a stellar feedback-driven origin. Across the full $M_\ast$ range of our sample, supernova energy coupling efficiencies of the order of $0.1{-}1\%$ are typically sufficient to account for the transformation from an initial NFW halo to the cored distribution we infer from the observations (Figure~\ref{fig:energy}). We therefore find no evidence for an energetic obstacle to forming the observed cores through stellar feedback.

\item The inner dark matter densities and logarithmic slopes of our sample are in fair agreement with those measured in the NIHAO, FIRE-2, and EDGE hydrodynamical simulations (Figure~\ref{fig:simulations_inner}). This agreement is found both at a fixed physical scale ($300\,\rm{pc}$) and at a homologous halo radius ($0.015\, R_{200}$). At these small scales, observations and simulations show similar departures from the collisionless NFW predictions at intermediate masses, consistent with feedback-induced core formation. The main residual discrepancy concerns the logarithmic slopes of some of the most massive simulated galaxies, which are steeper than those inferred from observations, possibly due to baryon-induced halo contraction or limitations in observational mass models.

\item At larger scales (5 kpc and $0.060\, R_{200}$), data and simulations follow NFW expectations more closely and broadly agree with each other (Figure~\ref{fig:simulations_outer}). Yet we find some differences at 5 kpc that can be traced to differences in the mapping between galaxies and haloes through the SHMR rather than to a systematic disagreement in the dark matter density profiles at fixed $M_{200}$. In fact, the simulations themselves follow somewhat different SHMRs, and the observed SHMR also appears broader than those of the simulations (Figure~\ref{fig:shmr}), although part of this latter difference may arise from substantial observational and mass modelling uncertainties.
\item Applying the $(\eta_{\rm bar},\eta_{\rm rot})$ diagnostic used in previous work, we find that our curated sample avoids the same regions of parameter space as NIHAO (Figure~\ref{fig:santos}). The galaxies responsible for much of the previously reported tension generally have substantial uncertainties in their kinematics or baryonic mass distributions, indicating that observational and modelling uncertainties are a major driver of the previously reported apparent discrepancy and its extreme tail.
\end{itemize}

Taken together, the results shown in this paper reveal substantial diversity in the inner dark matter density profiles of nearby galaxies, but no evidence of a systematic inner density mismatch relative to current $\Lambda$CDM hydrodynamical simulations. The energetic requirements for producing the inferred cores are modest, and the simulations fairly reproduce both the observed range and trends in inner dark matter densities and slopes. Within the observational and numerical limitations discussed in this paper, these findings substantially alleviate the cusp-core and diversity-of-rotation-curves problems.

A key next step will be to obtain homogeneous, high-quality SFHs for the galaxies in our sample and determine whether the scatter in their inner dark matter densities correlates with the duration and burstiness of their SFHs, as predicted by the simulations. Detecting these correlations would provide more direct support for feedback-driven core formation, whereas their absence would challenge this interpretation and strengthen the motivation for alternative dark matter models. In particular, self-interacting dark matter (SIDM; \citealt{spergel2000,tulin_sidm}) may be able to reproduce the range of rotation curve shapes observed in dwarf galaxies \citep{kamada2017,ren_SIDM_SPARC,zenter2022,nadler2023,roberts2025} without requiring a direct causal connection between SFHs and core formation. Warm dark matter (WDM; \citealt{colin2000,vladimir2001}) may offer a similar situation, since WDM haloes are less concentrated than their CDM counterparts, making it easier to form feedback-induced cores and requiring less bursty and less extended SFHs (e.g. \citealt{gonzalez2016}). In forthcoming work, we will compare the predictions of different alternative dark matter models with the kinematics of our galaxy sample. 

It will also be important to analyse large samples of gas-rich ultra-diffuse galaxies, which appear to have lower dark matter densities than expected for CDM \citep{huds2019,sengupta2019,shi2021,demao,agc114905,nadler2023,agc114905_deep}, and efforts are underway.\\

\begin{acknowledgements}
We thank Laura Sales, Gabriele Pezzulli, Kyle Oman, and Isabel Santos-Santos for valuable comments on our results. We also thank the FIRE and NIHAO collaborations for permission to use their public and private simulations, and Isabel Santos-Santos for sharing the NIHAO measurements obtained in \citetalias{santos2020}. We greatly appreciate the feedback from an anonymous referee, who helped clarify various aspects of our paper.

PEMP is funded by the Dutch Research Council (NWO) through the Veni grant VI.Veni.222.364. 
JSA thanks the Spanish Ministry of Economy and Competitiveness (MINECO) for support through grant P/301404 from the Severo Ochoa project CEX2019-000920-S. JSA acknowledges support by the grant PID2024-156100NB-C22 financed by MICIU/AEI/10.13039/501100011033/FEDER, EU. JSA thanks the Agencia Estatal de Investigación, CNS2023-144669, proyecto `TINY', and the 2024 call "Proyectos de Generación de Conocimiento", grant number PID2024-160009NA-I00, proyecto "INGENIO", PI A. DI Cintio.
CM is supported by a Sweden’s Wallenberg Academy Fellowship.

We have used SIMBAD \citep{simbad}, ALADIN \citep{aladin}, NED, and ADS extensively, as well as the tool TOPCAT \citep{topcat} and the Python packages NumPy \citep{numpy}, Matplotlib \citep{matplotlib}, SciPy \citep{scipy}, corner \citep{corner}, Colossus \citep{colossus}, and Astropy \citep{astropy}, for which we are thankful.
\end{acknowledgements}

   \bibliographystyle{aa.bst} 
   \bibliography{references} 

\begin{appendix}
\onecolumn
\section{The n parameter of the \textsc{coreNFW} profile}
\label{app:nparameter}
\begin{figure}[h]
    \begin{minipage}{0.52\linewidth}
        \includegraphics[width=\linewidth]{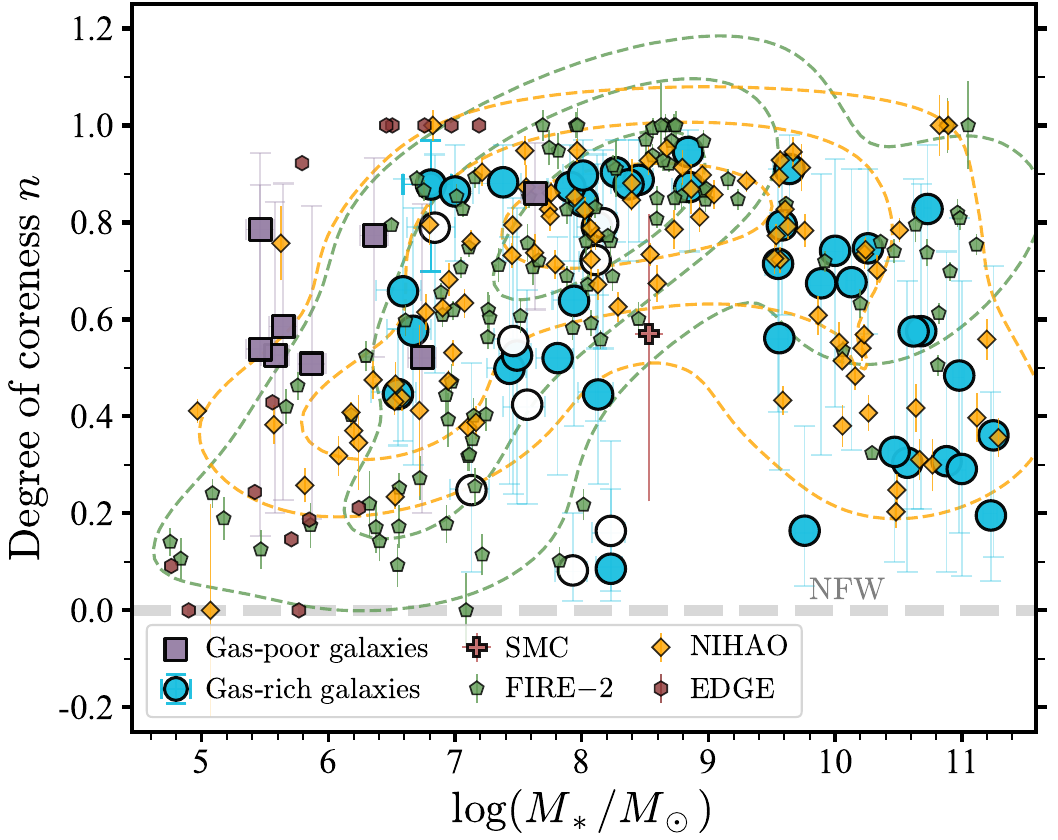}
    \end{minipage}%
    \hspace{0.03\linewidth}%
    \begin{minipage}{0.45\linewidth}
        \captionof{figure}{\textsc{coreNFW} $n$ parameter as a function of $M_\ast$ for our sample of galaxies and simulated systems. Symbols and contours are as in Figure~\ref{fig:simulations_inner}. The grey dashed line indicates $n\,{=}\,0$, i.e. cuspy NFW haloes.
        }
        \label{fig:simulations_ncorenfw}
    \end{minipage}
\end{figure}

As shown in Figure~\ref{fig:example_cNFW}, the parameter $n$ regulates the degree of coreness or the deviation of the \textsc{coreNFW} profile from an NFW, which shows substantial scatter at fixed $M_\ast$. In Figure~\ref{fig:simulations_ncorenfw} we compare the observational sample against EDGE, NIHAO, and FIRE-2, whose $n$ parameters are taken directly from \cite{muni2025} and \cite{sarrato2026}. Figure~\ref{fig:simulations_ncorenfw} shows that the hydrodynamical simulations also reproduce the overall trends and broad distribution in the data, although more observational constraints at the low-mass end are necessary to judge this agreement further.

\section{Comparison between data and simulations at larger scales and differences in SHMRs}
\label{app:large_scale}

Before this Appendix, we suggest the reader read Section~\ref{sec:simulations}, where we show good agreement between data and simulations for the density and density slopes at small scales. Here, we focus on the outer scales of 5 kpc and $0.060\, R_{200}$, which should probe regions less affected by baryonic effects. We note that observed gas-poor galaxies are excluded from this comparison, since for them the larger scales are either already in the region affected by tidally induced mass losses, or too far away from the centres, requiring a very large extrapolation compared to the extent of the stellar kinematic data (see \citealt{read2019,read2017_grav}). We will return to this issue in future work with improved data and analysis tools, but \cite{sarrato2026_dsph} find fair agreement between observed and simulated gas-poor satellites within the observational uncertainties.

\begin{figure}[h]
    \begin{minipage}{0.72\linewidth}
        \includegraphics[width=\linewidth]{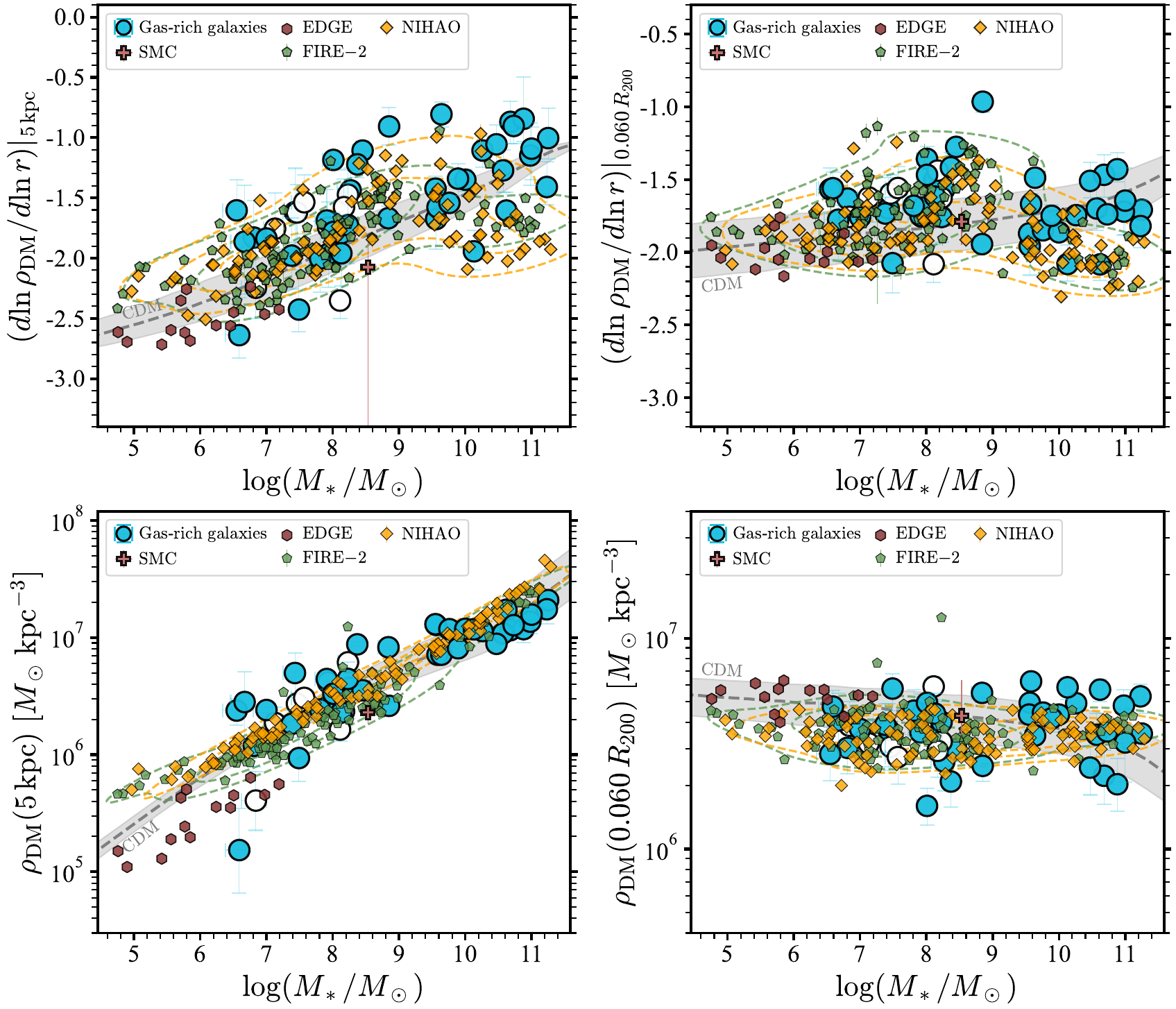}
    \end{minipage}%
    \hspace{0.03\linewidth}%
    \begin{minipage}{0.25\linewidth}
        \captionof{figure}{Dark matter logarithmic slope (top) and density (bottom) measured at 5 kpc (left) and $0.060\, R_{200}$ (right) for our observational sample and comparison simulations. Symbols are as in Figure~\ref{fig:simulations_inner}. White symbols indicate simulated systems with a resolution limit larger than the probed scale.}
        \label{fig:simulations_outer}
    \end{minipage}
\end{figure}

In Figure~\ref{fig:simulations_outer}, we inspect the slopes and densities at the outer scales of 5 kpc and $0.060\, R_{200}$ (Figures~\ref{fig:simulations_outer_m200} and \ref{fig:simulations_outer_fstar} present the same comparison but as a function of $M_{200}$ and $M_\ast/M_{200}$, respectively). The various panels of Figure~\ref{fig:simulations_outer} show that the agreement between data and simulations remains when probing larger radii. As expected, at these larger scales, the slopes become steeper, the densities lower, and both data and simulations lie closer to the CDM predictions, consistent with probing regions where dark matter heating is less critical.

\begin{figure}[h]
    \begin{minipage}{0.47\linewidth}
        \includegraphics[width=\linewidth]{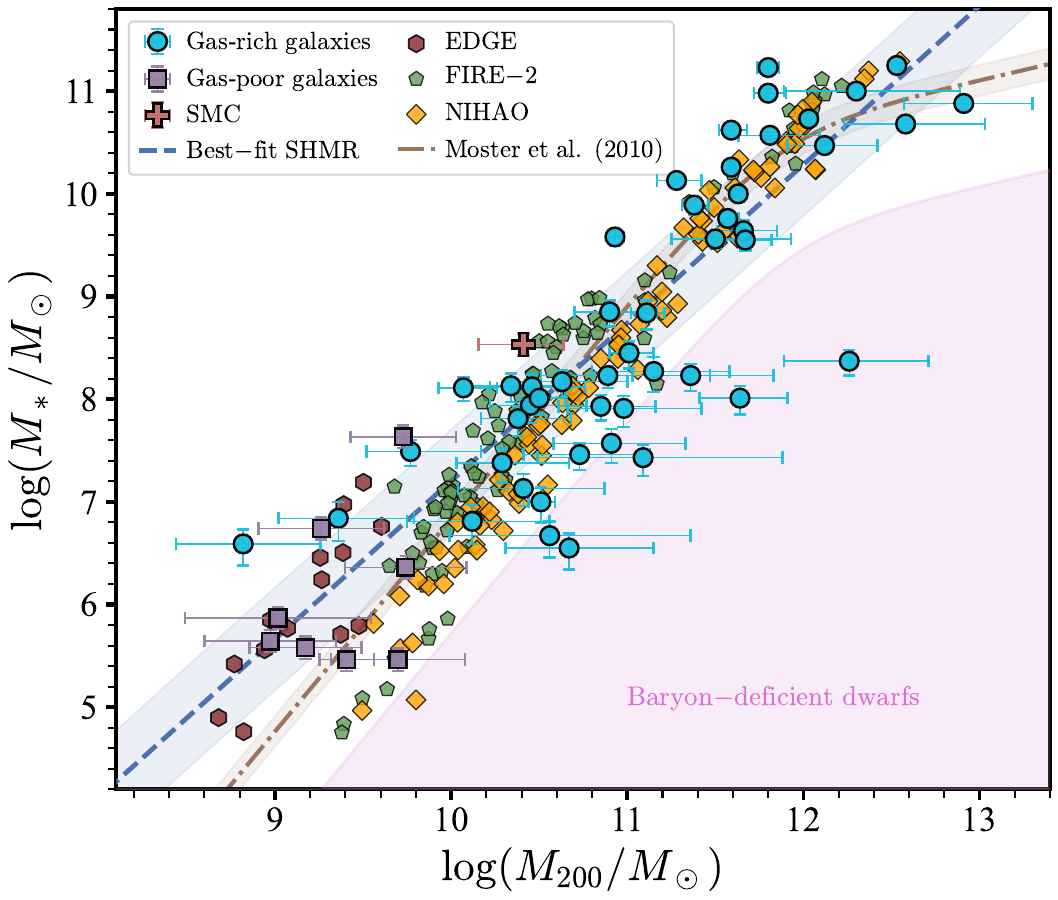}
    \end{minipage}%
    \hspace{0.03\linewidth}%
    \begin{minipage}{0.5\linewidth}
        \captionof{figure}{Stellar-to-halo mass relation (SHMR) for our observed and simulated galaxy samples. The dashed-dotted curve shows the SHMR from \cite{moster2010}, and the blue dashed line shows the best-fit relation to our observed sample: $\log(M_\ast/M_\odot)=1.54\,[\log(M_{200}/M_\odot)-11] + 8.74$, with a scatter of about 0.5 dex. The purple region represents the parameter space of `baryon-deficient' dwarfs (see \citetalias{paper_galaxyhalo}).
        }
        \label{fig:shmr}
    \end{minipage}
\end{figure}

In addition to the overall good agreement between the observed and simulated dark matter densities, Figure~\ref{fig:simulations_outer} reveals some subtle differences at 5 kpc. At the low-mass end, EDGE galaxies have lower densities than NIHAO and FIRE-2 galaxies, while NIHAO and FIRE-2 also show slightly different dependences of $\rho_{\mathrm{DM}}(5\,\mathrm{kpc})$ on $M_\ast$. A similar offset may extend to the observations, but there are too few directly comparable galaxies in this $M_\ast$ range to determine whether this is the case.
These differences follow directly from the galaxy--halo connection shown in Figure~\ref{fig:shmr}. Over their overlapping $M_\ast$ range, NIHAO galaxies generally inhabit more massive haloes at fixed $M_\ast$ than FIRE-2 galaxies, which in turn inhabit more massive haloes than EDGE galaxies. Their $\rho_{\mathrm{DM}}(5\,\mathrm{kpc})$ follow the same ordering because $\rho_{\mathrm{DM}}(5\,\mathrm{kpc})$ depends strongly on $M_{200}$. The offsets are much less apparent at $0.060\, R_{200}$ because this scale probes homologous regions of the haloes. Consistent with this interpretation, observations and simulations have compatible $\rho_{\mathrm{DM}}(5\,\mathrm{kpc})$ when compared at fixed $M_{200}$, as shown in Figure~\ref{fig:simulations_outer_m200}. The differences at fixed $M_\ast$ therefore primarily reflect differences in the SHMR rather than a systematic mismatch between the halo density profiles at fixed $M_{200}$.

Figure~\ref{fig:shmr} also shows that the inferred distribution of $M_{200}$ at fixed $M_\ast$ appears substantially broader in the observations than in the simulations. A similarly broad distribution was found in \citetalias{paper_galaxyhalo} when comparing observations with the TNG50 \citep{tng} and Simba \citep{simba} simulations.
As discussed in \citetalias{paper_galaxyhalo}, some but not all of the scatter can be attributed to a population of central `baryon-deficient' dwarfs whose $M_\ast$ is more than ten times lower than expected from the \cite{moster2010} relation (see also e.g. \citealt{forbes2024} for more examples of overly massive haloes). However, comparing observed and simulated dispersions directly is difficult because the former includes measurement and mass-modelling uncertainties, particularly those associated with extrapolating the fitted halo beyond the radial extent of the kinematic data, while halo masses are measured directly in the simulations. Yet, the overall match of the SHMR appears to remain an open issue in cosmological simulations, as it seems to be reproducing the stellar mass-size relation of nearby galaxies (e.g. \citealt{kaplinghat2020, sales_review_dwarfs,cruz2026}).

\vspace{-1cm}
\section{Slopes and densities as a function of $M_{200}$ and $M_\ast/M_{200}$}
\label{app:additional}
\begin{figure}[h]
    \begin{minipage}{0.72\linewidth}
        \includegraphics[width=\linewidth]{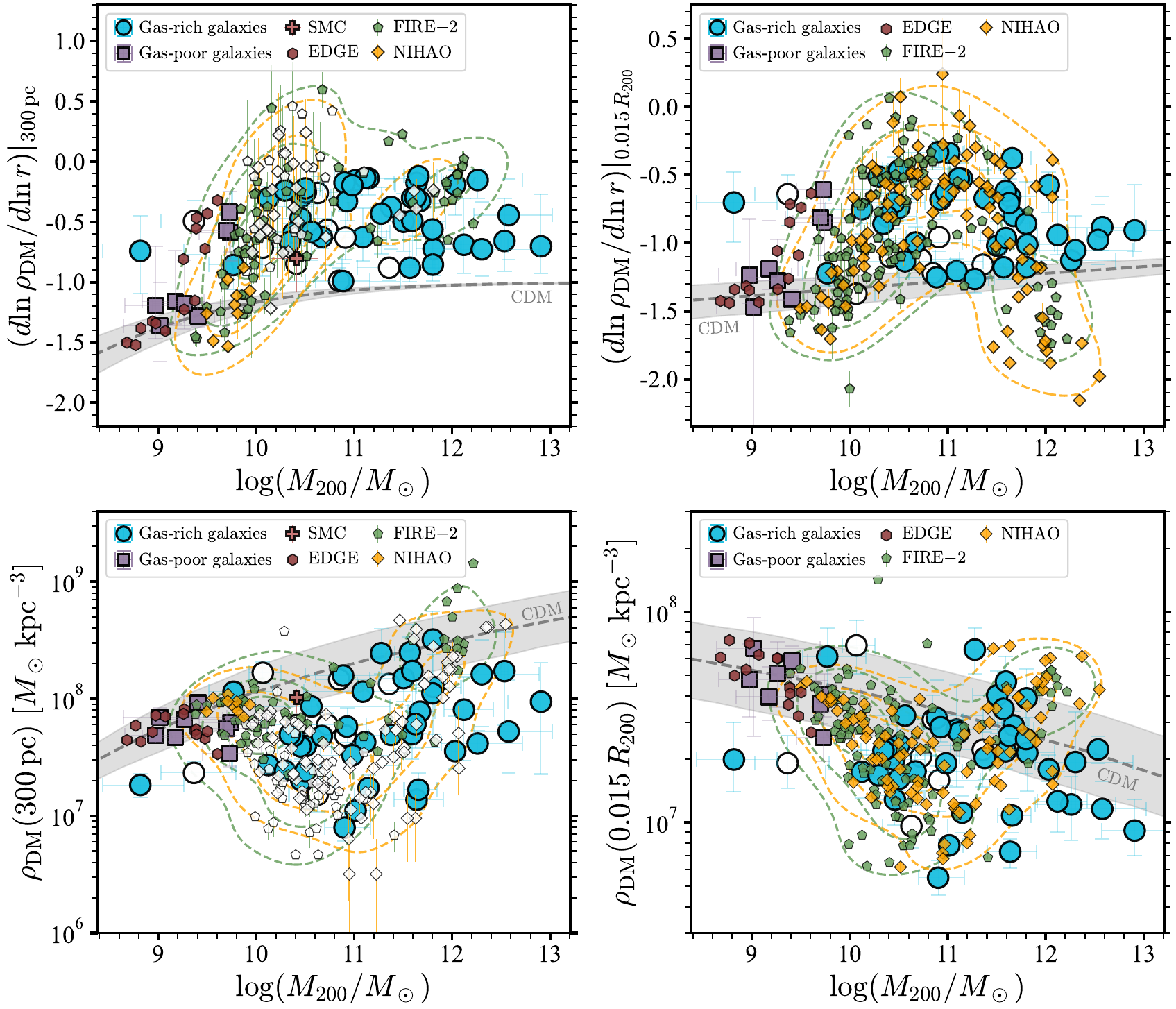}
    \end{minipage}%
    \hspace{0.03\linewidth}%
    \begin{minipage}{0.25\linewidth}
        \captionof{figure}{Same as Figure~\ref{fig:simulations_inner} but with slopes and densities as a function of $M_{200}$.
        }
        \label{fig:simulations_inner_m200}
    \end{minipage}
\end{figure}

\begin{figure}[h]
    \begin{minipage}{0.72\linewidth}
        \includegraphics[width=\linewidth]{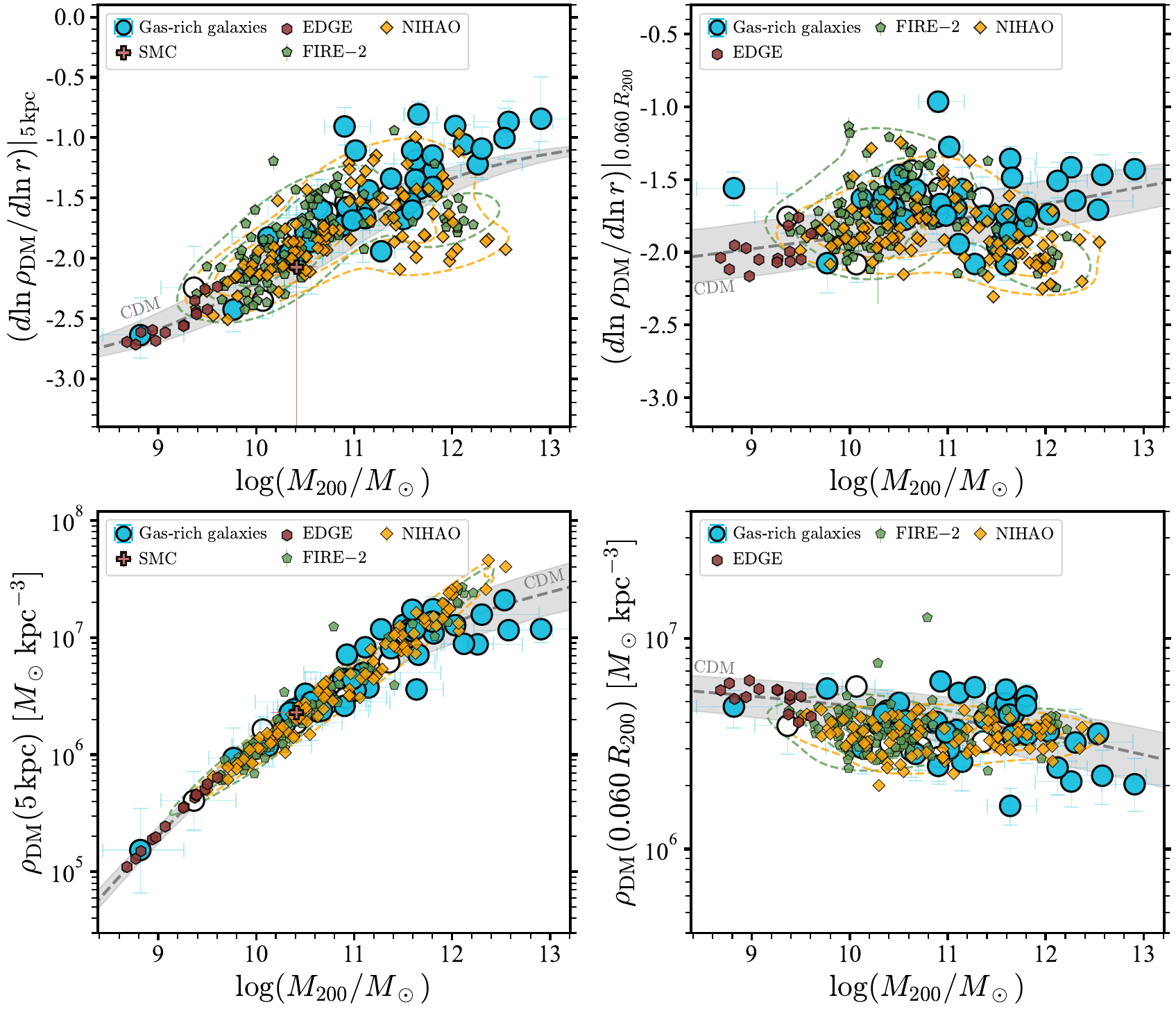}
    \end{minipage}%
    \hspace{0.03\linewidth}%
    \begin{minipage}{0.25\linewidth}
        \captionof{figure}{Same as Figure~\ref{fig:simulations_outer} but with slopes and densities as a function of $M_{200}$.\\
        }
        \label{fig:simulations_outer_m200}
    \end{minipage}
\end{figure}

\begin{figure}[!htbp]
    \begin{minipage}{0.72\linewidth}
        \includegraphics[width=\linewidth]{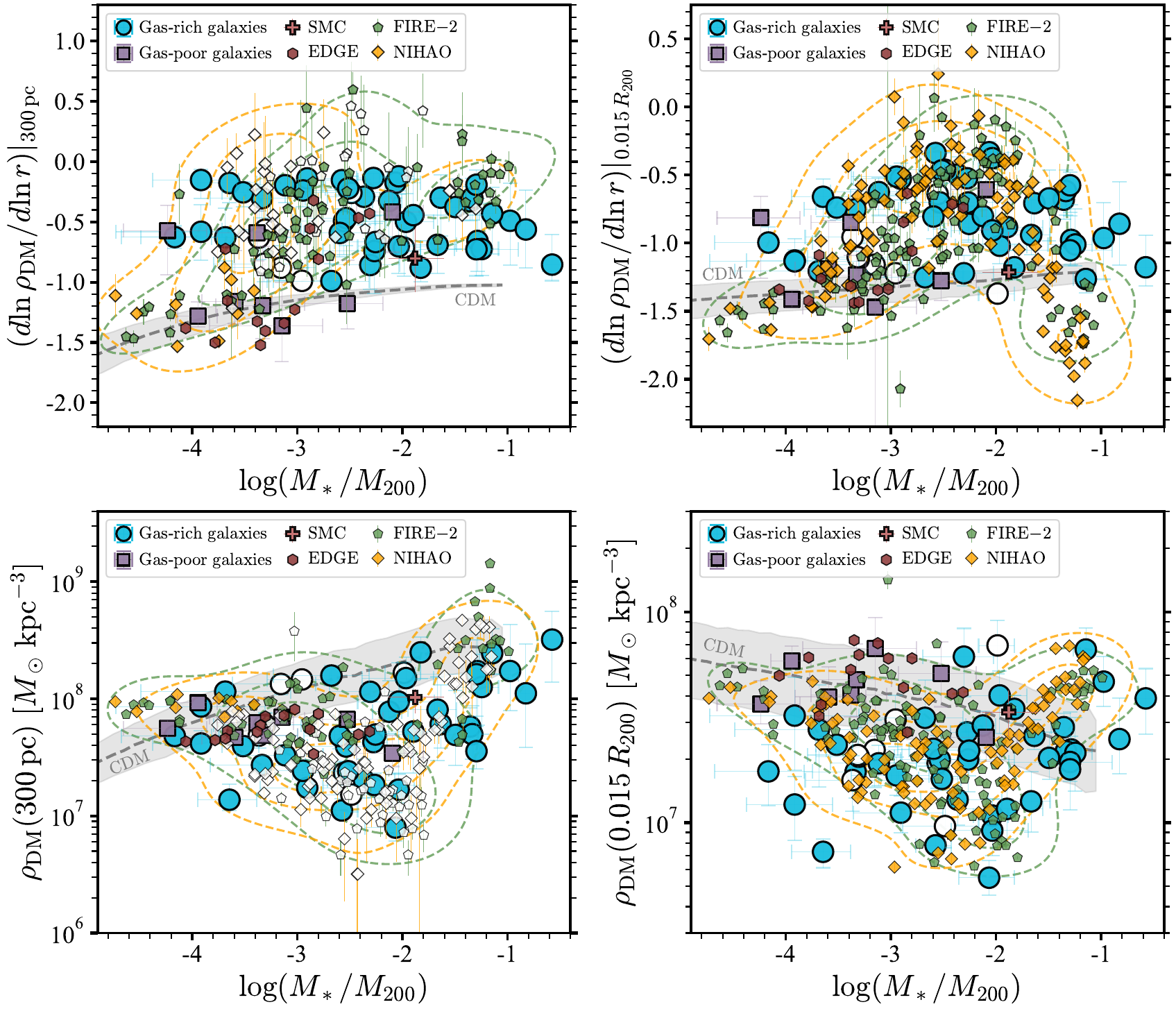}
    \end{minipage}%
    \hspace{0.03\linewidth}%
    \begin{minipage}{0.25\linewidth}
        \captionof{figure}{Same as Figure~\ref{fig:simulations_inner} but with slopes and densities as a function of $M_\ast/M_{200}$.
        }
        \label{fig:simulations_inner_fstar}
    \end{minipage}
\end{figure}

\begin{figure}[h]
    \begin{minipage}{0.72\linewidth}
        \includegraphics[width=\linewidth]{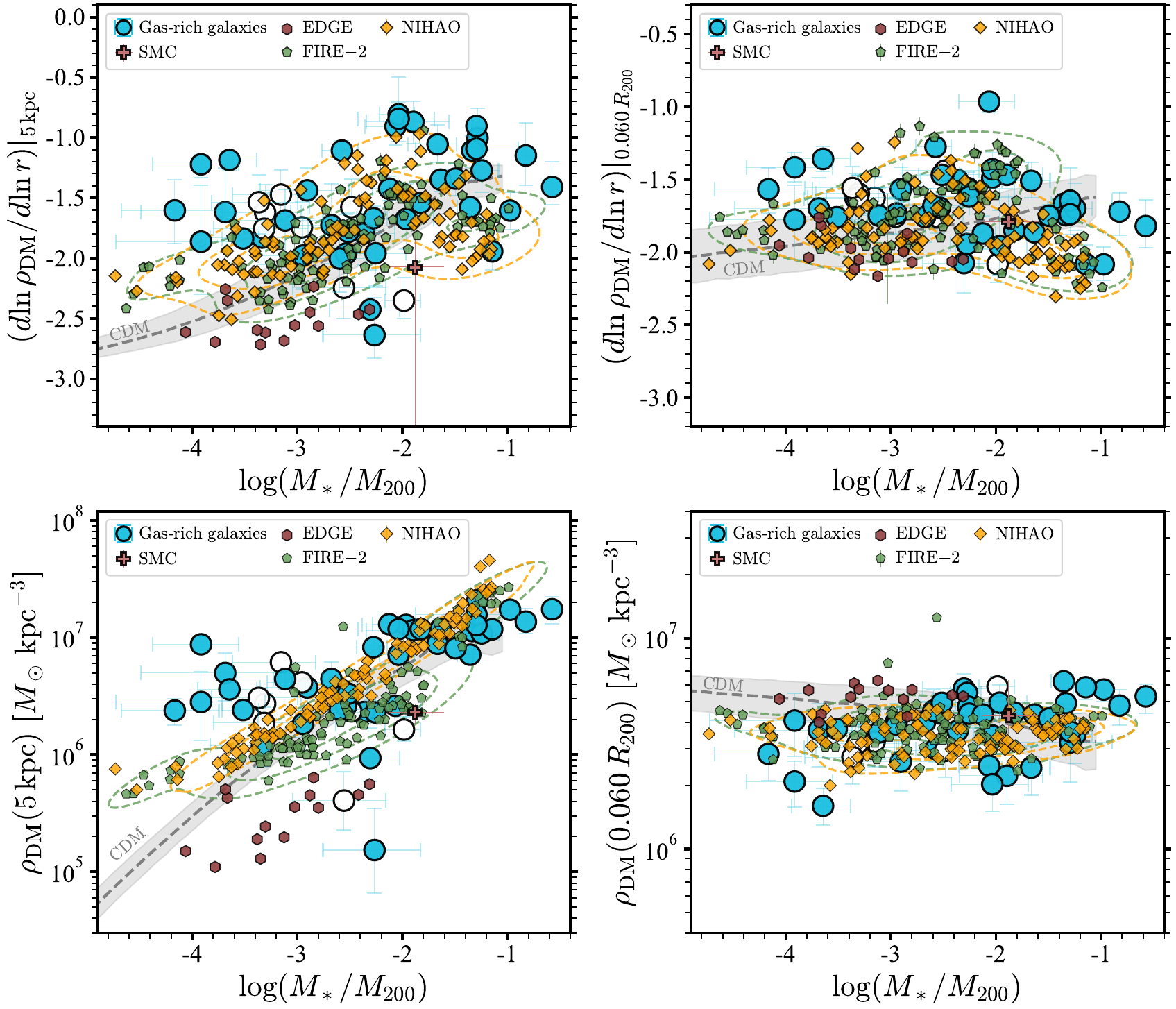}
    \end{minipage}%
    \hspace{0.03\linewidth}%
    \begin{minipage}{0.25\linewidth}
        \captionof{figure}{Same as Figure~\ref{fig:simulations_outer} but with slopes and densities as a function of $M_\ast/M_{200}$.
        }
        \label{fig:simulations_outer_fstar}
    \end{minipage}
\end{figure}

In Figures~\ref{fig:simulations_inner} and \ref{fig:simulations_outer}, we show the relation between different inner dark matter densities and slopes as a function of $M_\ast$. Here we present equivalent plots as a function of $M_{200}$ (Figures~\ref{fig:simulations_inner_m200} and \ref{fig:simulations_outer_m200}) and as a function of $M_\ast/M_{200}$ (Figures~\ref{fig:simulations_inner_fstar} and \ref{fig:simulations_outer_fstar}).

In Figure~\ref{fig:simulations_inner_m200}, broad agreement remains at both 300 pc and $0.015\,R_{200}$, although the trends are less clear when plotted against $M_{200}$ than against $M_\ast$. In the right panels, for instance, the characteristic mass dependence is weaker in the data than in the simulations, although some of this may reflect the uncertain slopes of the most massive galaxies, as discussed in connection with Figure~\ref{fig:simulations_inner}. The agreement in Figure~\ref{fig:simulations_outer_m200} remains fair and closer to the NFW-CDM reference. Figures~\ref{fig:simulations_inner_fstar} and \ref{fig:simulations_outer_fstar} present the largest scatter, as expected given the scatter in the SHMR: galaxies with similar $M_\ast/M_{200}$ but different $M_{200}$ can have very different dark matter densities at fixed physical scales. Also, because of the bend at high $M_\ast$ in the assumed SHMR \citep{moster2010}, the family of CDM haloes does not reach $M_\ast/M_{200}$ values as high as observed in some disc galaxies, as also reported in earlier work (e.g. \citealt{postinomissing,enrico_massmodels_ss}; \citetalias{paper_galaxyhalo}). Yet, in the context of the diversity-of-rotation-curves problem, the comparisons are encouraging, particularly in $\rho_{\rm DM}$.

\clearpage

\section{Impact of mass-to-light ratios on the cores of massive disc galaxies}
\label{app:massive}

Uncertainties in stellar mass-to-light ratios are usually not critical for low-mass galaxies, whose inner dynamics are dominated by dark matter and whose baryonic mass is often dominated by \hi. By contrast, $\Upsilon_{\rm d}$ and $\Upsilon_{\rm b}$ become crucial in massive galaxies, whose inner dynamics are often dominated by their stellar components and therefore suffer from the well-known disc--halo degeneracy \citep{vanalbada86}, further complicated in the presence of bulges. 

\begin{figure*}[!htbp]
    \centering
    \includegraphics[width=0.96\linewidth]{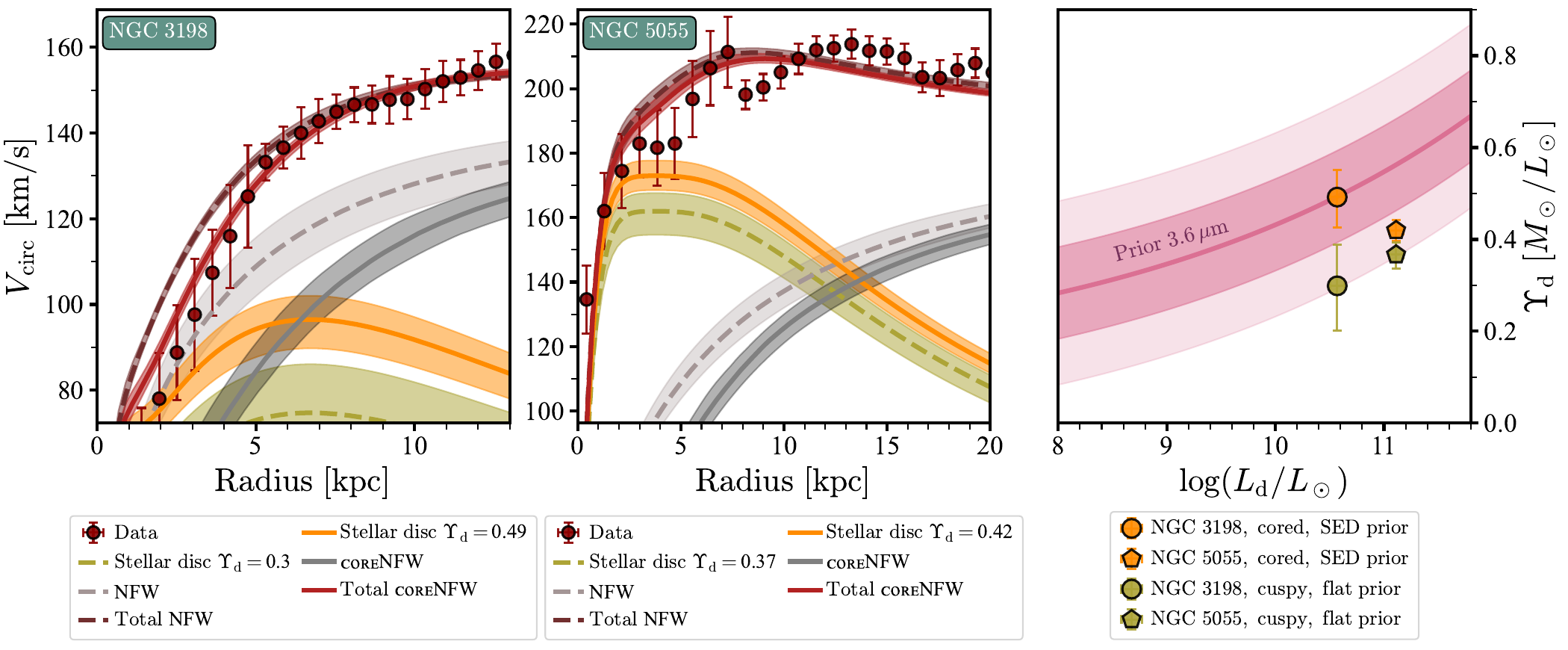}
    \caption{Left and middle: Mass models of two massive galaxies whose best-fit haloes have a relatively large $n$ parameter.
    Observed circular speeds are shown with red markers, while the circular speeds from the stellar disc, \textsc{coreNFW} halo, and total model are shown with the orange, grey, and red solid curves, respectively. The models also include \hi and H$_2$, but their contributions fall below our plotting range. 
    The fiducial models are compared against alternative models (dashed curves) that consider a cuspy NFW halo and a lower $\Upsilon_{\rm d}$ (after assuming a flat prior, instead of one informed by SED results, such as the fiducial model).
    Right: Comparison between the best-fitting $\Upsilon_{\rm d}$ and a $3.6\,\mu$m luminosity--$\Upsilon_{\rm d}$ relation from \citetalias{paper_galaxyhalo}, with shaded areas representing the $1\,\sigma$ and $2\,\sigma$ confidence bands. The relation was imposed as a Gaussian prior in the default (cored) mass models.
    }
    \label{fig:massive_cored}
\end{figure*}

Figure~\ref{fig:massive_cored} illustrates how this uncertainty affects the evidence for cores in two massive disc galaxies. To facilitate the comparison, the figure focuses on the radii and velocities near the centres. We compare our fiducial \textsc{coreNFW} models, which use an SED-informed prior on $\Upsilon_{\rm d}$, with deliberately conservative cuspy models with $n\,{=}\,0$ and for which $\Upsilon_{\rm d}$ is assigned a flat prior. The latter gives the NFW halo greater freedom to compensate for its central cusp by lowering the stellar contribution. Within the adopted mass models, NGC~3198 retains a preference for a core even when $\Upsilon_{\rm d}$ is allowed to vary freely: lowering $\Upsilon_{\rm d}$ is insufficient for the NFW model to reproduce the shape of the inner rotation curve as closely as the \textsc{coreNFW} model. For NGC~5055, however, a lower but still plausible $\Upsilon_{\rm d}$ substantially reduces the difference between the cored and cuspy fits. 

These examples suggest that modest cores may persist in some galaxies with $M_\ast\,{\gtrsim}\,10^{10}\, M_\odot$ (which have also been found in some simulations, e.g. \citealt{coreEinasto}), but also demonstrate that the strength of the evidence can be highly sensitive to the baryonic mass decomposition, which is even more aggravated for galaxies with bulges, both in terms of mass-to-light ratios and their kinematic modelling. Improved SED-fitting techniques and stellar-population models should provide tighter constraints on $\Upsilon_{\rm d}$ and $\Upsilon_{\rm b}$. It will also be important to explore the effects of radially varying mass-to-light ratios inferred through resolved SED fitting (e.g. \citealt{andreea,anastasia2026}).

\end{appendix}
\end{document}